\documentclass[10pt,aps,pra,twocolumn,amssymb,amsmath,%
    a4paper,twoside,superscriptaddress,showpacs,noshowkeys]{revtex4-1}

\usepackage{graphicx}

\renewcommand*{\Re}{\mathrm{Re}\,}
\renewcommand*{\Im}{\mathrm{Im}\,}

\DeclareMathOperator{\erfi}{erfi}
\DeclareMathOperator{\Ordo}{{\mathcal O}}
\newcommand*{\omegabar}{{\overline\omega}}
\newcommand*{\detuning}{{\delta}}
\newcommand*{\deltaomega}{{\Delta\omega}}
\newcommand*{\deltagap}{\Delta_{\text{gap}}}

\begin{document}
\title{Spectroscopic properties of inhomogeneously broadened spin ensembles in a cavity}
\author{Z. Kurucz}
\altaffiliation[Also at ]{Research Institute for Solid State Physics and Optics, H.A.S., H-1525 Budapest, Hungary}
\affiliation{Lundbeck Foundation Theoretical Center for Quantum System Research, Department of Physics and Astronomy, University of Aarhus, DK-8000 {\AA}rhus C, Denmark}

\author{J. H. Wesenberg}
\affiliation{Centre for Quantum Technologies, National University of Singapore, Singapore 117543}

\author{K. M{\o}lmer}
\affiliation{Lundbeck Foundation Theoretical Center for Quantum System Research, Department of Physics and Astronomy, University of Aarhus, DK-8000 {\AA}rhus C, Denmark}

\date{25 January, 2010}
\begin{abstract}
In large ensembles of identical atoms or spins, the interaction with a mode of the electromagnetic radiation field concentrates in a single superradiant degree of freedom with a collectively enhanced coupling. Given a controllable inhomogeneous broadening, such ensembles may be used for multi-mode storage of quantum states of the radiation field with applications in quantum communication networks and quantum computers. In this paper we analyze how the width and shape of the inhomogeneous broadening influence the collective enhancement and the dynamics of the cavity-ensemble system with focus on the consequences for the ensemble's applicability for quantum information processing tasks.

\end{abstract}
\pacs{%
  42.50.Ct, 
  42.50.Pq, 
  42.50.Gy
}%
\maketitle

\section{Introduction}
\label{introduction}

In ensembles of a large number of identical atoms, the interaction with the electromagnetic radiation field concentrates in a few collective degrees of freedom. Typically, one identifies effective oscillator degrees of freedom, which represent the collective atomic population of different internal states. The strong effective coupling of these oscillators to incident quantum fields makes atomic ensembles a prospective component in light-matter interfaces~\cite{RevModPhys.82.1041}, quantum memories~\cite{PhysRevLett.84.5094}, repeaters for long-range quantum communication~\cite{Nature.414.413}, and many other applications in quantum information technology. 

The ensemble size on the one hand provides a large optical depth, and on the other hand it provides phase matching conditions to couple strongly to weak quantum fields. In a microscopic quantum formulation, an incident single photon couples to a single collective excitation, i.e., to a superposition state where all atoms have the same or similar excitation amplitude. The corresponding Rabi frequency is proportional to the square root of the number of atoms. While a single collective oscillator degree of freedom is coupled to the field, a stored field state can be transferred to another spatial mode of collective excitation by applying a controlled reversible inhomogeneous broadening (CRIB) to the atoms. This way, a multimode optical interface and memory can be established \cite{OptLett.8.483}. Although optical transitions of rare earth ion dopants in crystals are inhomogeneously broadened, narrow spectral features can still be defined with hole burning techniques, and both CRIB schemes~\cite{OptCommun.247.393, PhysRevA.73.020302} and schemes based on atomic frequency combs~\cite{PhysRevA.79.052329} have been used to demonstrate storage of up to 1060 pulses of light~\cite{PhysRevLett.104.040503, arXiv.1009.2317}.

In this paper, we consider an ensemble of $N$ effective spin-$1/2$ particles, coupled to a central quantum oscillator (cavity). A cavity can be used to significantly enhance the atom-light interaction, as demonstrated by experiments on ultracold rubidium atoms~\cite{Nature.450.268, *Nature.450.272} and ion Coulomb crystals~\cite{NatPhys.5.494} in an optical cavity. Furthermore, a single quantized field mode can be used as an interface in hybrid proposals for quantum computing, e.g., to couple polar molecules \cite{NatPhys.2.636, PhysRevLett.101.040501} or solid state spin ensembles~\cite{PhysRevLett.103.070502} to a superconducting qubit via a transmission-line resonator. Recent experimental breakthroughs have led to the observation of strong collective coupling between a superconducting transmission-line resonator and large ensembles ($N>10^{12}$) of electron spins of chromium ions in ruby~\cite{PhysRevLett.105.140501} or nitrogen-vacancy centers in diamond~\cite{PhysRevLett.105.140502}, while iron nuclei embedded in a low-$Q$ planar cavity have been resonantly excited by synchrotron radiation to a superradiant state with a large collective Lamb shift~\cite{Science.328.1248}. The multimode capacity of ensembles has also been observed with nitrogen electron spins in fullerene cages and the electron and nuclear spins of phosphorous in silicon~\cite{PhysRevLett.105.140503}.

The purpose of this paper is to investigate collective enhancement in the presence of inhomogeneous broadening, which is not controllable by the experimentalist. We study a system with an interaction free Hamiltonian
\begin{gather}
  \label{eq:H0.spin}
  \hat H_0 = \omega_c \hat a_c^\dag \hat a_c + \sum_j \omega_j \hat \sigma_z^j,
\end{gather}
where $\omega_c$ is the angular frequency of the central oscillator, $\omega_j$ is the time-independent transition frequency of the $j$th spin, and $\hbar=1$.  The spins are non-interacting and the coupling to the central oscillator is described by a Jaynes--Cummings Hamiltonian,
\begin{gather}
  \label{eq:H1.spin}
  \hat H_1 = \sum_j g_j \hat \sigma^j_+ \hat a_c + \text{H.c.},
\end{gather}
with possibly different coupling constants $g_j$.

This model has fundamental importance and it appears with different variations in quantum physics: a single spin coupled to a bath of oscillators (the model explaining spontaneous emission of light from a single atom), a central spin coupled to a spin bath in a spin-star configuration \cite{PhysRevB.70.045323}, a single oscillator interacting with a spin ensemble, or a cloud of atoms in an optical or microwave cavity.  Many of these models have been addressed in textbooks \cite{cct-api, *peskin-qft} with emphasis on the resulting dissipative dynamics of the central system.  The mathematical difficulties arising when taking the limit of a continuously dense ensemble are also well-known \cite{PhysRev.95.1329, *CommunPureApplMath.1.361, caianiello61, OptCommun.179.247}.

Here we revisit the problem with emphasis on the ``spin bath'' degrees of freedom. We show that the density of spin states plays an important role in the joint dynamics of the central oscillator and the collective spin wave mode that is directly coupled to it. In certain cases, despite the inhomogeneity, all the other spin wave modes are effectively decoupled. This fact manifests itself in reduced oscillator linewidths determined by the homogeneous linewidth of the individual spins~\cite{PhysRevA.53.2711, Grenoble}. For other configurations, however, the oscillator linewidths are dominated by the inhomogeneous broadening, and the ``decoherence'' of the superradiant mode is collectively enhanced.

The paper is organized as follows.  Sec.~\ref{sec:dressed} derives some important properties of the model system, including the distribution of eigenenergies, the transmission spectrum through the cavity, and a formal solution for the dynamics of the system in the limit of high polarization. In Sec.~\ref{sec:e.g.}, we analyze in detail the cases where the effective density of spin states is Gaussian and Lorentzian, and compare in the two cases the oscillator linewidths, as well as the degree of Rabi splitting. In Sec.~\ref{sec:strong}, we consider the regime of strong coupling and study the protective effects of the Rabi splitting in a perturbative manner. Section~\ref{sec:conclusion} summarizes our results.

\section{Dressed ensemble}
\label{sec:dressed}

At low temperatures (high polarization), a collection of two-level systems and a collection of oscillators behave identically.  This fact is conveniently described in the Holstein--Primakoff approximation by introducing the bosonic operators $\hat a_j$ for each spin-$1/2$ particle
\begin{gather}
  \label{eq:HP}
  \hat \sigma_z^j \equiv -\tfrac12 + \hat a_j^\dag \hat a_j,
  \quad
  \hat \sigma_+^j \equiv \hat a_j^\dag \sqrt{1 - \hat a_j^\dag \hat a_j}
  \approx \hat a_j^\dag.
\end{gather}
The nonlinearity introduced by the square root term in Eq.~\eqref{eq:HP} ensures that no two excitations can take place at the same spin. If we consider delocalized spin waves involving a large number of spins compared to the number of excitations, the probability that a given spin is excited is inversely proportional to the number of spins $N$. Therefore, as long as only a few delocalized spin excitations are considered, it is reasonable to neglect the square root term in Eq.~\eqref{eq:HP}. In this regime, the free and interaction Hamiltonian \eqref{eq:H0.spin} and~\eqref{eq:H1.spin} become quadratic in the bosonic operators (apart from an omitted c-number),
\begin{gather}
  \label{eq:H0}
  \hat H_0 = \omega_c \hat a_c^\dag \hat a_c +
  \sum_{j=1}^N \omega_j \hat a_j^\dag \hat a_j,
  \allowdisplaybreaks[1]
  \\
  \label{eq:H1}
  \hat H_1 = \sum_{j=1}^N g_j \hat a_j^\dag \hat a_c + \text{H.c.}
  = \Omega \big( \hat b^\dag \hat a_c + \hat a_c^\dag \hat b \big),
\end{gather}
where $\hat b^\dag \equiv \sum_j \alpha_j \hat a_j^\dag$, defined by the normalized vector $\alpha_j = g_j / (\sum_k |g_k|^2)^{1/2}$, is the creation operator of a delocalized spin wave mode, the so-called \emph{superradiant} mode.  This is the concentrated degree of freedom which is coupled to the cavity with the collective coupling strength $\Omega = ({\sum_k |g_k|^2})^{1/2}$ that scales as $\sqrt N$.  An important consequence of the Holstein--Primakoff approximation is that the excitations become independent (noninteracting) quasiparticles.  Therefore, the dynamics of a single excitation provides the general solution, even if the total number of excitations is actually much larger than unity (but still much smaller than~$N$, see~\cite{PhysRevB.82.024413} for the case when the number of excitations in the superradiant mode is comparable to the number of spins).

The system consisting of a central oscillator interacting with a discrete or continuous bath of oscillators as described by Eqs.~\eqref{eq:H0} and~\eqref{eq:H1} has been studied in many physical contexts \cite{cct-api, *peskin-qft, PhysRev.95.1329, *CommunPureApplMath.1.361, caianiello61, OptCommun.179.247,PhysRevB.70.045323}. In the following, our aim is to gain a detailed understanding of the spectroscopic signature of the bath, as observed through the cavity mode.

In order to do so, we will in Sec.~\ref{sec:graphsol} describe the eigenbasis of $\hat H$ and in Sec.~\ref{sec:Green} use the resolvent formalism \cite{cct-api, *peskin-qft} to calculate the matrix elements of the time evolution operator. We find that the eigenbasis as well as the evolution operator is closely linked to the \emph{level-shift function}, and we devote Sec.~\ref{sec:analytic} to studying the analytic properties of this function before applying our analysis to describe transmission spectroscopy in Sec.~\ref{sec:transmission}.

\subsection{Exact solution of the eigenvalue problem for discrete systems}
\label{sec:graphsol}

In the high polarization (low excitation) limit, the total Hamiltonian $\hat H = \hat H_0 + \hat H_1$ is quadratic in the creation and annihilation operators, $\hat H = \sum_{\mu\nu} H_{\mu\nu} \hat a_\mu^\dag \hat a_\nu$ with $\mu$, $\nu = c$,~1, \ldots,~$N$. We introduce the ``dressed'' eigenmodes
\begin{gather}
\hat\Phi_q^\dag \equiv \eta_{qc} \hat a_c^\dag + \sum_j \eta_{qj} \hat a_j^\dag,
\end{gather}
that bring the total Hamiltonian into the diagonal form $\hat H = \sum_q E_q \hat \Phi_q^\dag \hat \Phi_q$.  For atomic spins in an optical cavity, the quasiparticles corresponding to the excitations of these eigenmodes are called cavity polaritons.  We will refer to $\eta_{qc}$ as the \emph{photonic amplitude} of the $q$th polariton mode, while $\eta_{qj}$ ($j=1$, \ldots,~$N$) is the \emph{mode function} of the spin wave part.  In field theory, $\eta_{qc}$ is referred to as the wave function renormalization constant \cite{OptCommun.179.247}.

In what follows, we will assume that the spins are nondegenerate.  Otherwise, if $m\ge2$ spins have the same transition frequency $\omega_j$, we can replace them with a single effective spin $\hat a_j' = \sum_k^m g_k \hat a_k /g_j'$ and an enhanced coupling constant $g_j' = (\sum_k^m |g_k|^2)^{1/2}$. The other $m-1$ orthogonal combinations are uncoupled.

\begin{figure}
  \centering
  \includegraphics[width=.9\hsize]{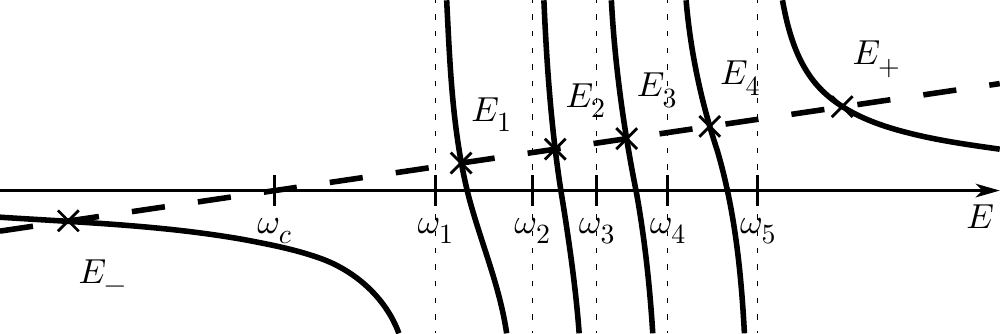}
  \caption{Schematic graph visualizing the solutions of
    Eq.~\eqref{eq:root-for-Eq}. The intersections of the rational
    function $\tilde K(E)$ (solid) with the $y=E-\omega_c$ line
    (dashed) give the eigenenergies~$E_q$.}
  \label{fig:root-for-Eq}
\end{figure}

The $N+1$ eigenenergies in the single-excitation subspace are the solutions for $E_q$ of the equation
\begin{gather}
  \label{eq:root-for-Eq}
  E_q-\omega_c - \tilde K(E_q) = 0,
  \end{gather}
where $\tilde K$ is the level-shift function~\cite{cct-api, *peskin-qft}
  \begin{gather}
  \label{eq:K(z).discrete}
  \tilde K(z) \equiv \sum_j \frac{|g_j|^2 }{z-\omega_j},
\end{gather}
which we will discuss in more detail in Sec.~\ref{sec:analytic}. The poles of the rational function $\tilde K(z)$ are located at the different values of $\omega_j$.  For real $\omega_\mu$-s, the solutions of Eq.~\eqref{eq:root-for-Eq} can be found graphically as the intersections of $\tilde K(E)$ with the line $y = E-\omega_c$ (see Fig.~\ref{fig:root-for-Eq}).
For most of the eigenvalues the perturbation only causes a small shift: Except for two modes, $E_-<\omega_{\text{min}}$ and $E_+>\omega_{\text{max}}$, all the perturbed eigenvalues are within the region $[\omega_{\text{min}}, \omega_{\text{max}}]$. In fact,  assuming increasingly ordered spin transition frequencies, $\omega_1 < \omega_2 < \ldots < \omega_{N}$, we see that  $\omega_q < E_q < \omega_{q+1}$ with $q=1,\ldots,N-1$.

\begin{figure*}
  \centering
  \centerline{\includegraphics[scale=.2222222222]{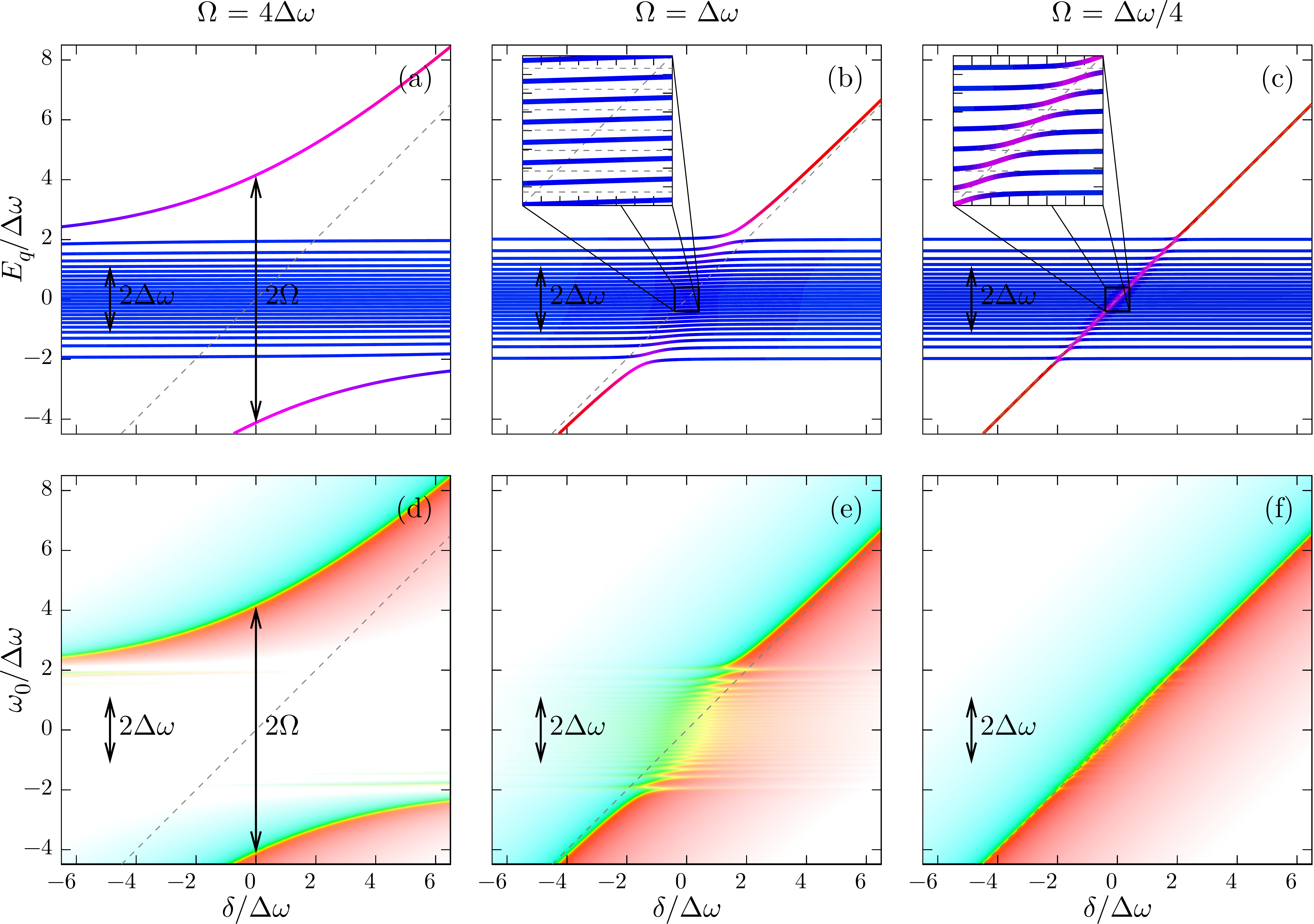}}
  \caption{(Color online) Numerical calculation for $N=25$ spins with an inhomogeneous width of $\deltaomega$, sampled from a Gaussian distribution. The individual spins decay at a rate $\gamma_j = 0.1 \deltaomega$ and are uniformly coupled to the cavity. The cavity decay rate is $\kappa = 0.05 \deltaomega$. (a)--(c) The real part of the exact eigenenergies as function of the cavity detuning.  Color red (blue) indicates that the corresponding polariton is photon-like (spin-like). (d)--(f) Transmission spectrogram, the color hue (saturation) corresponds to the phase (logarithmic magnitude) of the susceptibility \eqref{eq:suscept}. In the strong coupling regime, $\Omega = 4 \deltaomega$, (a) and~(d), only the two extremal eigenmodes have photonic attribute. In the regime of intermediate coupling, $\Omega = \deltaomega$, (b) and~(e), a band of intermediate eigenmodes has significant cavity content, and the cavity can ``decay'' into any of these modes. In the weak coupling regime, $\Omega = \deltaomega/4$, (c) and~(f), there is no collective effect, and only a few, resonant spins are dressed by the cavity.}
  \label{fig:crossing}
\end{figure*}

The photonic amplitude and the mode function of the normalized eigenmode $q$ can be obtained, respectively, as
\begin{gather}
  \label{eq:etaq}
  \eta_{qc}^{-2} = 1 + \sum_j
  \frac{|g_j|^2}{|E_q-\omega_j|^2},
  \qquad
  \eta_{qj} = \frac{g_j \eta_{qc}}{E_q-\omega_j}.
\end{gather}
As an example, Figs.~\ref{fig:crossing}a--c show the eigenenergies and the photonic amplitudes of the eigenmodes for an ensemble of 25 spins in the regimes of strong, intermediate, and weak coupling.

\subsection{Matrix elements of the time evolution operator}
\label{sec:Green}

We will now calculate the transition amplitude for an excitation created at $t=0$ in the oscillator mode $\nu$ to end up in oscillator $\mu$ at a later time $t$, that is, we calculate the matrix elements of the evolution operator $\hat U(t) \equiv \exp {(-i \hat H t)}$ on the single-excitation subspace.
These transition amplitudes are also called Green's functions and can be expressed as commutator expectation values in the Heisenberg picture,
\begin{gather}
  \label{eq:Gmunu.green}
  G_{\mu\nu}(t) \equiv \left[ e^{-i\mathbf Ht} \right]_{\mu\nu} = \big\langle \big[ \hat a_\mu(t),
  \hat a_\nu^\dag(0) \big] \big\rangle.
\end{gather}
To account for photons leaking out of the cavity at a rate $\kappa$ and spins decaying at rate $\gamma_j$, we start from the Heisenberg--Langevin equation of motion for the annihilation operators in the absence of external driving field,
\begin{gather}
  \label{eq:ddt-ac}
  \frac d{dt} \hat a_c = -i \omega_c \hat a_c
  - i \sum_j g_j^* \hat a_j + \hat f_c,
  \\
  \label{eq:ddt-aj}
  \frac d{dt} \hat a_j = -i \omega_j \hat a_j - i g_j \hat a_c + \hat f_j,
\end{gather}
where $\hat f_c$ and $\hat f_j$ are Langevin noise operators satisfying $\langle f_c(t) f_c^\dag(t') \rangle = 2\kappa \delta({t-t'})$ and $\langle f_j(t) f_k^\dag(t') \rangle = \gamma_j \delta_{jk} \delta({t-t'})$, while the cavity frequency and the spin transition frequencies are considered complex with negative or zero imaginary parts, $\Im \omega_c = -\kappa$ and $\Im \omega_j = -\frac12 \gamma_j$.  For the Green's functions \eqref{eq:Gmunu.green} we obtain a system of differential equations,
\begin{gather}
  \label{eq:ddt-Gcnu}
  i\frac d{dt} G_{c\nu}(t) = \omega_c G_{c\nu}(t) + \sum_j g_j^* G_{j\nu}(t),
  \\
  \label{eq:ddt-Gjnu}
  i\frac d{dt} G_{j\nu}(t) = \omega_j G_{j\nu}(t) + g_j G_{c\nu}(t),
\end{gather}
with the initial condition $G_{\mu\nu}(0)=\delta_{\mu\nu}$. Substituting into Eq.~\eqref{eq:ddt-Gcnu} the formal solution
\begin{gather}
  \label{eq:Gjnu(t)}
  G_{jc}(t) = G_{jc}(0) e^{-i\omega_jt}
  - ig_j \int_0^t G_{cc}(\tau) e^{-i\omega_j(t-\tau)} \, d\tau,
\end{gather}
we get a closed integro-differential equation for $G_{cc}(t)$:
\begin{gather}
  \label{eq:integro-diff}
  i\frac d{dt} G_{cc}(t) = \omega_c G_{cc}(t)
  -i \int_0^t G_{cc}(\tau) K(t-\tau) \, d\tau,
\end{gather}
where we have introduced the memory kernel function
\begin{gather}
  \label{eq:K(t)}
  K(t) \equiv \sum_j |g_j|^2 e^{-i\omega_jt}.
\end{gather}

Equation \eqref{eq:integro-diff} can be solved in the Fourier
domain by extending the integral to a proper convolution. For this, we introduce the
advanced ($+$) and retarded ($-$) versions of the Green's functions and
the memory kernel function,
\begin{gather}
  G^{\pm}_{\mu\nu}(t) \equiv \mp i \Theta(\pm t) G_{\mu\nu}(t)
  ,\nonumber \\
  \label{eq:def-pm}
  K^{\pm}(t) \equiv \mp i \Theta(\pm t) K(t),
\end{gather}
where $\Theta(t)$ is the Heaviside step function. With these functions, we have a proper convolution,
\begin{gather}
  \label{eq:integro-diff-pm}
  i\frac d{dt} G^{\pm}_{cc}(t) = \delta(t) + \omega_c G^{\pm}_{cc}(t)
 + \int_{-\infty}^\infty G^{\pm}_{cc}(\tau) K^{\pm}(t-\tau) \, d\tau.
\end{gather}
Now taking the Fourier transform of Eq.~\eqref{eq:integro-diff-pm} we obtain
\begin{gather}
  \label{eq:integro-diff-pm.fourier}
  \omega \tilde G^{\pm}_{cc}(\omega) = 1
  + \omega_c \tilde G^{\pm}_{cc}(\omega)
  + \tilde G^{\pm}_{cc}(\omega) \tilde K^{\pm}(\omega),
\end{gather}
from which the explicit form of the reduced dynamics of the central oscillator follows by simple algebra,
\begin{gather}
  \label{eq:Gcc-pm.fourier}
  \tilde G^{\pm}_{cc}(\omega) = \big[
  {\omega - \omega_c - \tilde K^{\pm}(\omega)}
  \big]^{-1}.
\end{gather}
The Fourier transforms $\tilde G^{\pm}_{\mu\nu}(\omega) = \int G^{\pm}_{\mu\nu}(t) e^{i\omega t} \,dt$ are conventionally called forward and backward propagators. We note, however, that the backward propagators $\tilde G_{\mu\nu}^- (\omega)$ are undefined if any of the spin relaxation rates $\gamma_j$ or the cavity $\kappa$ is non-zero. 

The remaining propagator matrix elements can all be expressed in terms of $\tilde G^\pm_{cc}(\omega)$: Introducing $k_j(t)\equiv e^{-i\omega_jt}$ and defining $k_j^{\pm}(t)$ similarly to Eq.~\eqref{eq:def-pm}, we find by Eq.~\eqref{eq:Gjnu(t)} that
\begin{gather}
  \label{eq:Gjc-pm.fourier}
  \tilde G^{\pm}_{jc}(\omega)
  = g_j \tilde k_j^{\pm}(\omega)
  \tilde G^{\pm}_{cc}(\omega),
\\
  \label{eq:Gck-pm.fourier}
  \tilde G^{\pm}_{ck}(\omega)
  = g_k^* \tilde k_k^{\pm}(\omega)
  \tilde G^{\pm}_{cc}(\omega),
  \\
  \label{eq:Gjk-pm.fourier}
  \tilde G^{\pm}_{jk}(\omega)
  = \delta_{jk} \tilde k_j^{\pm}(\omega)
  + g_j g_k^* \tilde k_j^{\pm}(\omega) \tilde k_k^{\pm}(\omega)
  \tilde G^{\pm}_{cc}(\omega).
\end{gather}
Furthermore, for the amplitude of transiting from the cavity to the superradiant spin wave mode, $G_{sc}(t) \equiv \big\langle \big[ \hat b(t), \hat a_c^\dag(0) \big] \big\rangle$, we have
\begin{gather}
  \label{eq:Gsc-pm.fourier}
  \tilde G_{sc}^\pm(\omega) = \tilde G_{cs}^\pm(\omega) =
  \Omega^{-1} \tilde K^\pm(\omega) \tilde G_{cc}^\pm(\omega) ,
  \\
  \label{eq:Gss-pm.fourier}
  \tilde G^{\pm}_{ss}(\omega)
  = 1 - \tilde K^{\pm}(\omega)
  \tilde G^{\pm}_{cc}(\omega).
\end{gather}

\subsection{Analytic properties of the level-shift function}
\label{sec:analytic}

We shall see that the complex analytic extension of the level shift $\tilde K^\pm(\omega)$ is of special importance~\cite{cct-api, *peskin-qft}. Here we summarize some of its properties. First we note that the Fourier--Laplace transforms $\tilde k_j^{\pm}(z) \equiv \int k_j^\pm(t) e^{izt} \, dt$ are defined on complementary halves of the complex plane,
\begin{gather}
  \label{eq:k-j-pm.fourier-laplace}
  \tilde k_j^\pm (z) =
  \begin{cases}
    \dfrac1{z-\omega_j}, & \mbox{if $\Im z \gtrless \Im\omega_j$,}\\
    \mbox{undefined},   & \mbox{if $\Im z \lessgtr \Im\omega_j$,}\\
  \end{cases}
\end{gather}
and when $\Im z$ approaches $\Im\omega_j = -\frac12 \gamma_j$ from above or below, $\tilde k_j^+ (z)$ and $\tilde k_j^- (z)$ tend to different distributions,
\begin{gather}
  \label{eq:k-j-pm.branchcut}
  \lim_{\eta\to0^+} \tilde k_j^\pm (\omega - \tfrac i2 \gamma_j \pm i\eta)
  = \mathcal P \frac1{\omega-\Re\omega_j}
  \mp i\pi\delta({\omega-\Re\omega_j}),
\end{gather}
where $\omega$ is real and $\mathcal P$ denotes the principal value.

\begin{figure}
  \centering
  \includegraphics{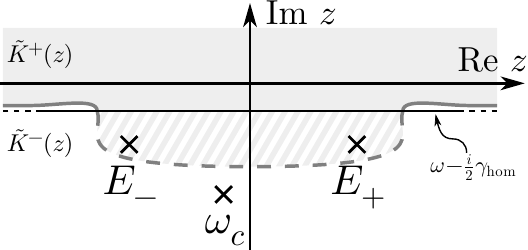}
  \caption{
    The branch cut of the level-shift function $\tilde K(z)$ displayed
    as the straight line ${\omega - \tfrac i2 \gamma_{\text{hom}}}$
    ($\omega \in \mathbb R$) in the complex plane. $\tilde K(z)$ has a
    jump of $\tilde K^-(z) - \tilde K^+(z) = i\Gamma_c(z)$ when
    passing through the cut from above.  Depending on whether $\kappa$
    is smaller than, equal to, or larger than $\tfrac 12
    \gamma_{\text{hom}}$, the poles of the cavity-cavity propagator,
    i.e., the roots $E_\pm$ of Eq.~\eqref{eq:resonance-peaks}, may lie
    above the cut, on the cut, or on the second Riemann sheet below
    the cut, respectively.
  }
  \label{fig:branch-cut}
\end{figure}

The whole ensemble may consist of discrete and continuous sets of spins, and $\tilde K^+ (z) = \sum_j |g_j|^2 \tilde k_j^+ (z)$ is defined for $\Im z > \sup_j \Im \omega_j$, while $\tilde K^- (z)$ is defined for $\Im z < \inf_j \Im \omega_j$. For simplicity, we assume that the decay rate is the same for all spins, $\gamma_j = \gamma_{\text{hom}}$. If the coupling density profile $\rho(\omega) \equiv \sum_j |g_j|^2 \delta ({\omega - \Re\omega_j})$ is an analytic function of the real variable $\omega$, then
\begin{gather}
  \label{eq:K.analytic}
  \tilde K(z) = \int \frac{\rho(\omega)}
  {z - \omega + \frac i2 \gamma_{\text{hom}}} \, d\omega
\end{gather}
is a complex analytic extension of $\tilde K^\pm (z)$, that has pole singularities in the discrete points $\omega_j$ and a branch cut discontinuity along the line ${\omega - \tfrac i2 \gamma_{\text{hom}}}$, such that it approaches different values from above and from below,
\begin{gather}
  \label{eq:K.branch-cut}
  \lim_{\eta\to0^+} \tilde K \big(\omega -\tfrac i2 \gamma_{\text{hom}} \pm i\eta \big)
  = \tilde K^\pm \big({\omega - \tfrac i2 \gamma_{\text{hom}}}\big),
\end{gather}
as illustrated in Fig.~\ref{fig:branch-cut}.
The real and imaginary parts of $\tilde K^\pm (z) = \Delta_c (z) \mp \tfrac i2 \Gamma_c (z)$ on the branch cut line $z={\omega-\tfrac i2\gamma_{\text{hom}}}$ follow from Eqs.~\eqref{eq:k-j-pm.branchcut} and~\eqref{eq:K.analytic},
\begin{gather}
  \label{eq:Deltac}
  \Delta_c \big(\omega - \tfrac i2 \gamma_{\text{hom}}\big) =
  \mathcal P \int \frac {\rho(\omega')} {\omega - \omega'} \,d\omega',\\
  \label{eq:Gammac}
  \Gamma_c \big({\omega - \tfrac i2 \gamma_{\text{hom}}}\big)
  = 2\pi \rho(\omega).
\end{gather}
Thus the analytic continuation of $\tilde K^+(z)$ is given by Eq.~\eqref{eq:K.analytic} above the branch cut, while it explores the second Riemann sheet of $\tilde K(z)$ below the cut,
\begin{gather}
  \label{eq:branch-cut-jump}
  \tilde K^+(z) =
  \begin{cases}
    \tilde K(z)& \mbox{if $\Im z > -\tfrac 12 \gamma_{\text{hom}}$,}\\
    \Delta_c (z) - \tfrac i2 \Gamma_c (z)
    & \mbox{if $\Im z = -\tfrac 12 \gamma_{\text{hom}}$,}\\
    \tilde K(z) - i\Gamma_c(z)
    & \mbox{if $\Im z < -\tfrac 12 \gamma_{\text{hom}}$.}\\
  \end{cases}
\end{gather}

We can revert Eq.~\eqref{eq:Gammac} and extract information about the coupling density from the values of the level-shift function $\tilde K^+(\omega)$ which is evaluated on the real axis,
\begin{multline}
  \label{eq:rho(omega).inverted}
  \rho(\omega) = -\frac1\pi \Im \tilde K^+
  \big(\omega - \tfrac i2 \gamma_{\text{hom}}\big)
  \\
  \approx -\frac1\pi \Im \tilde K^+ (\omega) + \frac 1{2\pi}
  \frac{\partial \Re \tilde K^+ (\omega)}{\partial \omega} \gamma_{\text{hom}},
\end{multline}
where we truncated the Taylor series of $\Im\tilde K^+(z)$ at first order in $\gamma_{\text{hom}}$ and obtained the derivative from the Cauchy--Riemann equations.

Finally, we remark that we included $\gamma_{\text{hom}}$ in Eq.~\eqref{eq:K.analytic} merely in order to separate the effects of homogeneous and inhomogeneous broadening. Alternatively, we could have assumed a coupling density profile that already includes homogeneous broadening via the convolution
\begin{gather}
  \label{eq:rho'(omega)}
  \rho'(\omega) \equiv \int \frac{\gamma_{\text{hom}}/{2\pi}}
  {(\omega-\omega')^2 + \tfrac14 \gamma_{\text{hom}}^2}
  \rho(\omega') \, d\omega'.
\end{gather}
The branch cut of the new level-shift function $\tilde K'(z) = \int {\rho'(\omega)}/({z-\omega}) \, d\omega$ is then on the real axis. However, the analytic continuations of $\tilde K'^\pm(z)$ are the same as those of $\tilde K^\pm(z)$, and the two models lead to the same dynamics.

\subsection{Linear response and spectroscopic features}
\label{sec:transmission}

An experimental tool for studying the eigenenergies of the coupled cavity-ensemble system is \emph{transmission spectroscopy}. In this process, the (lossy) cavity is driven by an external classical field of frequency $\omega_0$ via the interaction Hamiltonian $\hat V(t) = \mathcal E e^{-i\omega_0t} \hat a_c^\dag + \mbox{H.c}$. The transmitted field---which is proportional to the steady state of the field inside the cavity---reflects the linear response of the system to the driving field and shows a resonance when the driving frequency is near an eigenfrequency. Alternatively, one may consider, e.g., directly driving the spins via an external field that couples to the total spin, $\hat V(t) = \mathcal E e^{-i\omega_0t} \sum_j\hat a_j^\dag + \mbox{H.c}$.

Let us consider a general external classical drive described by $\hat V(t) = \sum_\mu \mathcal E_\mu e^{-i\omega_0t} \hat a_\mu^\dag + \mbox{H.c}$. We assume that the driving field is weak, so that the total number of excitations is always small and the Holstein--Primakoff approximation remains valid.  Then the Heisenberg--Langevin equation of motion for the annihilation operators can be obtained from Eqs.~\eqref{eq:ddt-ac} and~\eqref{eq:ddt-aj} by adding $-i\mathcal E_c e^{-i\omega_0t}$ and $-i\mathcal E_j e^{-i\omega_0t}$, respectively, to the right-hand sides. The formal solution of this set of inhomogeneous differential equations can be written in terms of the Green's functions as
\begin{multline}
    \label{eq:a(t).driven}
  \hat a_\mu (t) = \sum_\nu G_{\mu\nu}(t-t_0) \hat a_\nu(t_0)
  + \sum_\nu \int_{t_0}^t G_{\mu\nu}(t-\tau)
  \\\times
  \big[ -i \mathcal E_\nu e^{-i\omega_0\tau} + \hat f_\nu(\tau) \big] \, d\tau.
\end{multline}
The first term vanishes when taking the limit $t_0 \to -\infty$, for all eigenvalues of $H_{\mu\nu}$ are assumed to have negative imaginary part (even if infinitesimal). Since $\langle f_\mu(t) \rangle = 0$, the expectation value of the oscillator operators in the steady state becomes
\begin{gather}
  \big\langle a_\mu^{\text{ss}}(t) \big\rangle
  = e^{-i\omega_0t} \sum_\nu  \chi_{\mu\nu}(\omega_0) \mathcal E_\nu,
\end{gather}
where
\begin{gather}
  \label{eq:suscept}
  \chi_{\mu\nu}(\omega) \equiv
  -i \int_0^\infty G_{\mu\nu}(\tau) e^{i\omega\tau} \,d\tau
  = \tilde G_{\mu\nu}^+(\omega)
\end{gather}
is the susceptibility (impedance) of the system, which has been calculated in Eqs.\ \eqref{eq:Gcc-pm.fourier}--\eqref{eq:Gss-pm.fourier}.

In the transmission spectroscopic setup, the cavity field is $\big\langle a_c^{\text{ss}} (t) \big\rangle = \mathcal E e^{-i\omega_0t} \chi_{cc}(\omega_0)$, so the transmissivity of the system is proportional to $|\tilde G^+_{cc}(\omega_0)|^2$, and the phase shift is $\phi(\omega_0) = \arg\tilde G^+_{cc}(\omega_0)$ \cite{Grenoble, PhysRevA.77.063833}. The resonance peaks in the transmission spectrum are characterized by the complex poles of $\tilde G^+_{cc}(z)$, that is, the solutions for $z$ of the implicit equation
\begin{gather}
  \label{eq:resonance-peaks}
  \big[ \tilde G_{cc}^+ (z)\big]^{-1} =
  {z - \omega_c} - \tilde K^+(z) = 0.
\end{gather}
For discrete systems, the analytic continuation of $\tilde K^+(z)$ has pole singularities in the points $\omega_j$ in the lower half of the complex plane, and the solutions of Eq.~\eqref{eq:resonance-peaks} coincide with the eigenenergies $E_q$ obtained form Eq.~\eqref{eq:root-for-Eq}. For continuous systems described in Sec.~\ref{sec:analytic}, the roots of Eq.~\eqref{eq:resonance-peaks} may be located above the branch cut of $\tilde K(z)$, on the cut line, or below it depending on whether $\kappa$ or $\tfrac12 \gamma_{\text{hom}}$ is larger (see Fig.~\ref{fig:branch-cut}).

Finally, we mention that there is a direct way to extract information about the coupling density profile from a measured transmission spectrum. First the level-shift function is to be calculated by inverting Eq.~\eqref{eq:Gcc-pm.fourier}: $\tilde K^+(\omega) = \omega - \omega_c - \chi^{-1}_{cc}(\omega)$. Alternatively, if only the transmissivity is available experimentally but not the phase of the susceptibility,  $\tilde K^+(\omega)$ may be obtained by fitting a Lorentzian to the transmissivity $|\chi_{cc}(\omega,\omega_c)|^2$ as function of the cavity frequency $\omega_c$ (with $\Im\omega_c = -i\kappa$). Once the values of the level-shift function $\tilde K^+(\omega)$ are known for real frequencies, Eq.~\eqref{eq:rho(omega).inverted} can be used to derive the coupling density.

\section{Examples of spectroscopic signatures}
\label{sec:e.g.}

With the results of the previous section, we are equipped to predict the spectroscopic signature for a given system as observed by transmission spectroscopy. While the output signal could have been obtained more directly via classical input-output theory \cite{Grenoble}, we shall see that the framework developed in Sec.~\ref{sec:dressed} provides additional insights into the physics by explicitly considering the state of the spin bath. We will first consider the limiting cases of infinitely narrow and infinitely wide coupling density profiles, before studying the qualitative differences between the intermediate cases for Lorentzian and Gaussian coupling density profiles. 

\subsection{Oscillation and decay}
\label{sec:oscil-vs-decay}

In certain situations, the detailed structure of the coupling density profile is not important. On time scales much smaller than the inverse inhomogeneous width of the ensemble, that is, when the inhomogeneous broadening cannot be resolved, the ensemble behaves as a single oscillator with a collectively enhanced coupling constant.  In this case, the memory kernel function \eqref{eq:K(t)} can be approximated by $K(t) \approx \Omega^2 e^{-i\omega_a t}$, or equivalently, $\tilde K^\pm(\omega) = \Delta_c(\omega) \mp \tfrac i2 \Gamma_c(\omega)$ with $\Delta_c(\omega) \approx \mathcal P \Omega^2/(\omega-\omega_a)$ and $\Gamma_c(\omega) \approx 2\pi \Omega^2 \delta(\omega-\omega_a)$, where the ensemble's mean transition frequency $\omega_a$ is assumed to be real. The inverse Fourier--Laplace transform of Eqs.~\eqref{eq:Gcc-pm.fourier} and~\eqref{eq:Gsc-pm.fourier} show that the cavity and the superradiant mode undergo Rabi oscillation at the Rabi frequency $\Omega_R = \sqrt{ \Omega^2 + \detuning^2/4}$,
\begin{gather}
  \label{eq:Gcc.Rabi-limit}
  G_{cc}(t) = e^{-i(\omega_a + \detuning/2)t}
  \big( \cos \Omega_R t
  - i \cos\theta \sin \Omega_R t \big),\\
  \label{eq:Gsc.Rabi-limit}
  G_{sc}(t) =  -i e^{-i(\omega_a + \detuning/2)t}
  \sin\theta \sin \Omega_R t,
\end{gather}
where the detuning of the cavity, $\detuning = \omega_c - \omega_a$, and the mixing angle $\tan\theta \equiv 2\Omega/\detuning$ are assumed to be real.

In the opposite limit, when the inhomogeneous width is much larger than the collective coupling $\Omega$, and the spin transition frequencies are sufficiently dense to form a continuum, we are in the Weisskopf--Wigner regime. In analogy with the spontaneous emission of an excited atom, the cavity excitation effectively decays into the continuum of spin waves in this regime~\cite{PhysRevLett.103.163603}. The key assumption in the Weisskopf--Wigner approximation is that the kernel function $K(t)$ decays fast compared to the slowly varying envelope $G_{cc}(t) e^{i\omega_ct}$, and thus the frequency dependence of $\tilde K^\pm(\omega)$ can be neglected. The inverse Fourier--Laplace transform of Eq.~\eqref{eq:Gcc-pm.fourier} then gives an exponential decay,
\begin{gather}
  \label{eq:WW-solution}
  G_{cc}(t) = e^{-i[\omega_c + \Delta_c(\omega_c) - \tfrac i2 \Gamma_c(\omega_c)] t}.
\end{gather}
Here we see that $\Delta_c(\omega_c)$ is the shift of the cavity frequency and $\tfrac12 \Gamma_c(\omega_c)$ is an additional cavity decay rate due to the coupling to the dense spin reservoir.

Finally, we consider the general case when an initially photonic excitation decays into the continuum of spin wave modes, and we specify the asymptotic energy distribution of the excited spin. If $G_{cc}(t)$ converges to zero, it is either because the initially photonic excitation leaks out of the cavity, or because the photon is irreversibly converted into a spin-like excitation. From Eq.~\eqref{eq:Gjnu(t)} we have
\begin{gather}
  \label{eq:Gjc.asymptotic}
  \lim_{t\to\infty} G_{jc}(t) e^{i\omega_jt} =
  g_j \tilde G_{cc}^+(\omega_j),
\end{gather}
so the asymptotic probability that the $j$th spin is excited equals $\lim_{t\to\infty}|G_{jc}(t)|^2 = |g_j|^2 |\tilde G_{cc}^+(\omega_j)|^2$, assuming that $\gamma_j=0$. Then the asymptotic energy distribution of the excited spin, i.e., the probability that the transition frequency of the excited spin is $\omega$, reads
\begin{multline}
  \label{eq:p(omega)}
  p(\omega) \equiv \sum_j \delta (\omega-\omega_j) |G_{jc}(\infty)|^2
  = \rho(\omega) |\tilde G_{cc}^+(\omega)|^2
  \\
  = \frac1{2\pi} \frac{\Gamma_c(\omega)}
  {[\omega - \Re\omega_c - \Delta_c(\omega)]^2
    + [2\kappa + \Gamma_c(\omega)]^2/4}.
\end{multline}
In particular, it is proportional to a Lorentzian in the Weisskopf--Wigner regime, in analogy to the energy distribution of a photon spontaneously emitted by an atom.

\subsection{Lorentzian coupling density profile}
\label{sec:lorentzian}

When the cavity is coupled to a single atomic oscillator (e.g., a degenerate ensemble of homogeneously broad but identical atoms), we recover the well-known problem of a driven two-level atom or that of two coupled oscillators. In this simple case, the results of Sec.~\ref{sec:graphsol} can be directly applied. With an atomic transition frequency $\omega_a$, natural atomic linewidth $\gamma$, cavity detuning from the atomic transition $\detuning$, and cavity linewidth $2\kappa$, we take the complex frequencies $\omega_1 = {\omega_a -\tfrac i2 \gamma}$ and $\omega_c = {\omega_a + \detuning - i\kappa}$. The eigenenergies obtained from Eq.~\eqref{eq:root-for-Eq} are
\begin{gather}
  \label{eq:E.Rabi}
  E_\pm 
  = \omega_a -\tfrac i2 \gamma + \Omega_R \cos\theta \pm \Omega_R,
\end{gather}
where
\begin{gather}
  \label{eq:Rabi-freq-complex}
  \Omega_R \equiv \sqrt{\Omega^2
    + \big[\detuning -i \big(\kappa-\tfrac12\gamma\big)\big]^2/4}
\end{gather}
is the (complex) Rabi frequency, and for the (complex) mixing angle we have
\begin{gather}
  \label{eq:mixing-angle-complex}
  \cos\theta \equiv \frac {\detuning - i (\kappa-\tfrac12\gamma)}{2\Omega_R},
  \qquad
  \sin\theta \equiv \frac \Omega{\Omega_R}.
\end{gather}
In analogy with the dressed states of a driven two-level atom, the eigenmodes are $\hat\Phi_\pm^\dag = \eta_{\pm c} \hat a_c^\dag + \eta_{\pm s} \hat b^\dag$ with the photonic amplitudes
\begin{gather}
 \eta_{+c} = \frac{\cos(\theta/2)}{\sqrt{\cosh(\Im\theta)}},
 \qquad
 \eta_{-c} = \frac{\sin(\theta/2)}{\sqrt{\cosh(\Im\theta)}},
\end{gather}
and the superradiant amplitudes $\eta_{\pm s} = \pm \eta_{\mp c}$.

The memory kernel function defined in Eq.~\eqref{eq:K(t)} is $K(t) = \Omega^2 e^{-i\omega_a t - \gamma t/2}$ for such a single collective spin excitation, which yields $\tilde K^+ (\omega) = \Omega^2 / ({\omega - \omega_a +\tfrac i2 \gamma})$. The same memory kernel function is obtained for a large, inhomogeneously broadened ensemble with a Lorentzian coupling density profile. This is the case, for example, when each spin in the ensemble is uniformly coupled to the cavity with the same coupling constant $g = \Omega / \sqrt N$, and the spin transition frequencies are distributed according to
\begin{gather}
  \label{eq:Breit-Wigner}
  \mathcal D(\omega) = \frac{1}{2\pi}
  \frac{N\gamma}{(\omega-\omega_a)^2 + \gamma^2/4},
\end{gather}
where now $\gamma$ is the ensemble's inhomogeneous width (full-width at half-maximum) and the individual spins do not decay, and the coupling density profile is $\rho(\omega) = g^2 \mathcal D(\omega)$. All the dynamical properties of both the cavity and the superradiant spin wave mode are determined by $\tilde K^+ (\omega)$ alone and, thus, they are the same for a single (homogeneously broad) atomic oscillator and for an inhomogeneous ensemble with Lorentzian coupling density. Therefore, it cannot be determined from the linear response of the cavity whether the broadening is of homogeneous or inhomogeneous origin.

The eigenenergies \eqref{eq:E.Rabi} are the two complex poles of the forward propagator \eqref{eq:Gcc-pm.fourier},
\begin{gather}
  \label{eq:Gcc-pm.fourier.lorentz}
  \tilde G_{cc}^+ (\omega) = \frac{\omega-\omega_a+\tfrac i2 \gamma}
  {(\omega - E_+)(\omega - E_-)},
\end{gather}
corresponding to two peaks in the transmission spectrum:  The position and the width of the peaks are determined by the real and imaginary parts of $E_\pm$, respectively. The system undergoes damped Rabi oscillations, as can be seen, e.g., by taking the inverse Fourier transform of Eqs.~\eqref{eq:Gcc-pm.fourier.lorentz} and~\eqref{eq:Gsc-pm.fourier},
\begin{align}
  G_{cc}(t) &= e^{-i(\omega_a + \detuning/2)t - (\gamma/2+\kappa) t/2}
  \nonumber\\&\quad\times
  \left( \cos\Omega_Rt
    - i \frac{\cos\Re\theta}{\cosh\Im\theta} \sin\Omega_Rt \right),
  \label{eq:Gcc.Rabi-1atom}
  \\
  G_{sc}(t) &= -i e^{-i(\omega_a + \detuning/2)t - (\gamma/2+\kappa) t/2}
  \nonumber\\&\quad\times
  \big(\sin\Re\theta \sin\Omega_Rt
    -\tanh\Im\theta \cos\Omega_Rt \big).
  \label{eq:Gsc.Rabi-1atom}
\end{align}

\begin{figure}
  \centering
  \includegraphics[scale=.4]{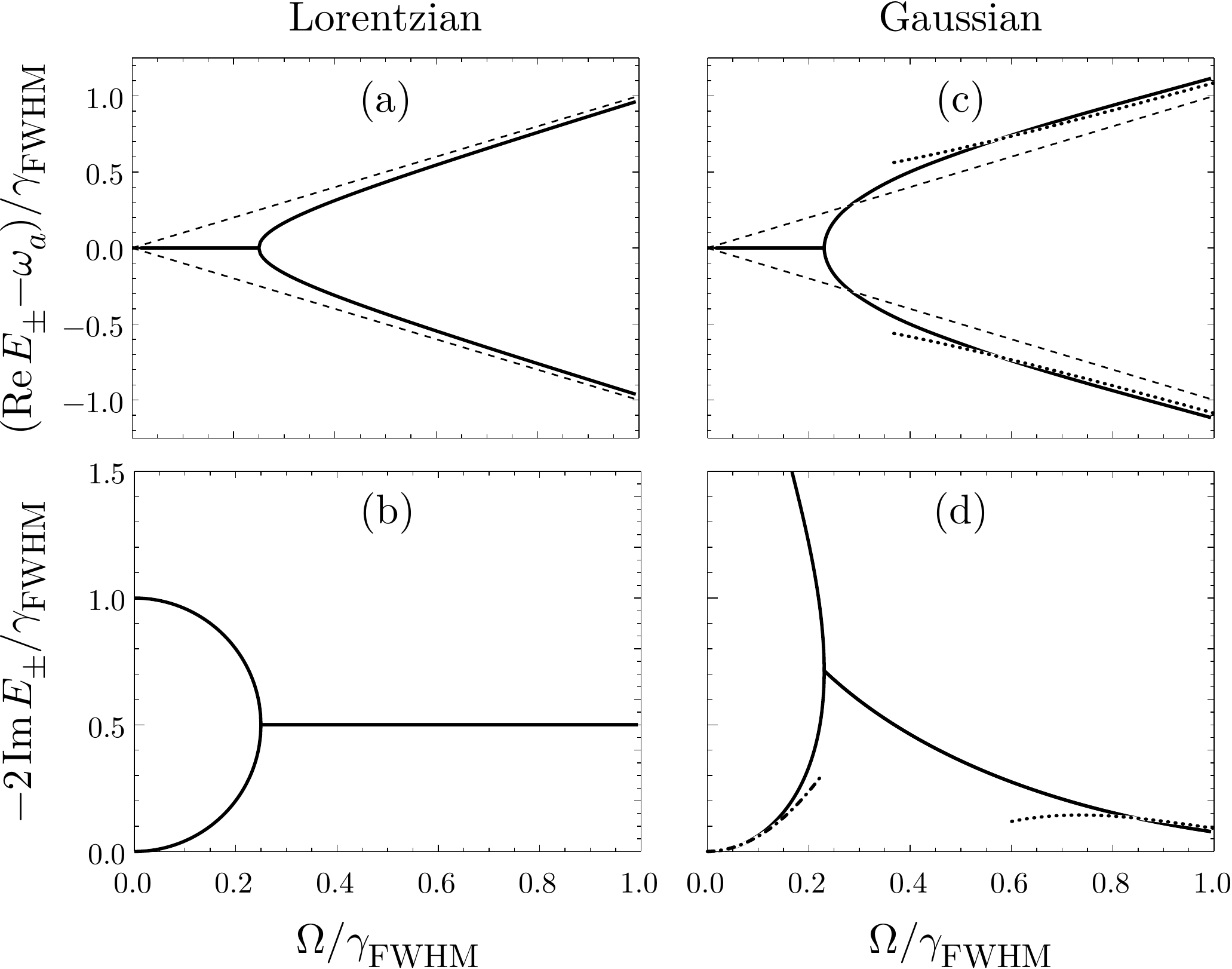}
  \caption{
    The real (a,~c) and imaginary (b,~d) parts of the poles of the
    susceptibility \eqref{eq:Gcc-pm.fourier} as function of the
    collective coupling strength for a Lorentzian (a,~b) and a
    Gaussian (c,~d) coupling density profile. The quantities are
    directly related to the position and width of the peaks in the
    transmission spectrum. The cavity is tuned at the center of the
    ensemble ($\omega_c=\omega_a$), and $\kappa = \gamma_{\text{hom}}
    = 0$ was assumed. The full-width at half-maximum of the coupling
    density is $\gamma_{\text{FWHM}}=\gamma$ for the Lorentzian and
    $\gamma_{\text{FWHM}}=\sqrt{8\ln2}\,\sigma$ for the Gaussian.  The
    dashed lines (a,~c) show the lowest order solutions of the
    asymptotic expansion given by Eq.~\eqref{eq:E0}, and correspond to
    the strong coupling approximation described in
    Sec.~\ref{sec:oscil-vs-decay}. The dotted curves (c,~d) show the
    asymptotic solutions in the next orders as given by
    Eqs.~\eqref{eq:ReE.gauss.1} and ~\eqref{eq:M2ImE.gauss.2}. The
    dash-dotted curve (c) shows Eq.~\eqref{eq:ww.gauss} corresponding
    to the Weisskopf--Wigner approximation in the weak coupling regime.
  }
  \label{fig:peaks}
\end{figure}

As an illustrative example, let us consider the resonant case $\detuning=0$.  Depending on the value of $(\gamma-2\kappa)/(4\Omega)$,  we see a transition from (i) damped Rabi oscillations at a reduced effective Rabi frequency to (ii) an overdamped situation (see also Figs.~\ref{fig:peaks}a--b).

\paragraph*{(i) Regime of oscillations.} The Rabi frequency \eqref{eq:Rabi-freq-complex} is real when ${\gamma-2\kappa} < 4\Omega$, giving rise to a Rabi splitting of $\Omega_R = \sqrt{ \Omega^2 - (\tfrac12 \gamma - \kappa)^2/4}$ around the atomic transition frequency $\omega_a$ (Fig.~\ref{fig:peaks}a), and the width of the peaks in the transmission spectrum is $-2\Im E_\pm = {\tfrac12 \gamma+\kappa}$ (Fig.~\ref{fig:peaks}b).  Here we see two important facts about a Lorentzian coupling density profile: the resonant Rabi splitting is smaller than the actual collective coupling strength $\Omega$, and the linewidth of the peaks is determined by the inhomogeneous width of the ensemble, independent of the coupling strength. As we shall see, the cavity does not provide protection for the superradiant mode in this case.

\paragraph*{(ii) Overdamped regime.} The Rabi frequency \eqref{eq:Rabi-freq-complex} is pure imaginary for ${\gamma-2\kappa} > 4\Omega$. There is no Rabi splitting at all, for we have two overlapping peaks at $\omega_a$ of width $-2\Im E_\pm = (\gamma + 2\kappa)/2 \mp \sqrt{ (\gamma - 2\kappa)^2/4 - 4\Omega^2}$.  In the Weisskopf--Wigner regime ($\gamma \gg \Omega, 2\kappa$), the dynamics reduces to the exponential decay given by Eq.~\eqref{eq:WW-solution}, and the peaks in the spectrum become clearly distinguishable: On top of a wide background peak with $\gamma_+ \approx \gamma$ corresponding to the weakly perturbed superradiant mode, there is superimposed a narrow peak of the cavity, whose width $\gamma_- \approx 2\kappa + 4\Omega^2 / \gamma$ equals the overall decay rate of the cavity excitation as given in Eq.~\eqref{eq:WW-solution}.

Finally we note that the oscillating--decaying dynamics of our cavity photon interacting with an inhomogeneously broadened, highly polarized spin ensemble shows some similarity with the dynamics of a single atom in a damped, zero temperature cavity~\cite{PhysRevA.29.2627}. In the latter case, the atom interacts with the lossy cavity, and the cavity itself is coupled to a collection of external radiation modes. In our case, the cavity interacts with the superradiant spin wave mode, and the superradiant mode itself is coupled to a collection of subradiant spin wave modes.

\subsection{Gaussian coupling density profile}
\label{sec:gaussian}

We consider now a Gaussian  coupling density profile,
\begin{gather}
  \label{eq:rho.gaussian}
  \rho(\omega) = \frac1{\sqrt{2\pi}}\frac{\Omega^2}{\sigma}
  \exp\left[-\frac{(\omega-\omega_a)^2}{2\sigma^2}\right],
\end{gather}
 and show that the spectrum of an ensemble with such an inhomogeneity is qualitatively different from what we learned from the Lorentzian profile in the previous section. From the integral in Eq.~\eqref{eq:K.analytic} we obtain
\begin{gather}
  \label{eq:K+.gaussian}
  \tilde K^+(z) = \sqrt{\frac\pi2} \frac{\Omega^2}{\sigma} e^{-\xi^2}
  \big[ \erfi(\xi) - i \big],
\end{gather}
where we have introduced the dimensionless parameter $\xi \equiv {(z-\omega_a+\tfrac i2 \gamma_{\text{hom}})} / (\sqrt2 \sigma)$.  Equation \eqref{eq:K+.gaussian} is composed of complex analytic functions with no branch cut singularities, and $\Im \tilde K^+({\omega - \tfrac i2 \gamma_{\text{hom}}}) = -\pi \rho(\omega)$. Therefore, Eq.~\eqref{eq:K+.gaussian} is indeed the analytic continuation of $\tilde K^+(z)$ on the entire complex plane. To determine the position and width of the peaks in the transmission spectrum, we solve Eq.~\eqref{eq:resonance-peaks} numerically (Figs.~\ref{fig:peaks}c--d). Here we observe two regimes: an overdamped one without Rabi splitting ($\Omega/\gamma_{\text{FWHM}} \lesssim 0.23$) and another one with two Rabi-split peaks ($\Omega/\gamma_{\text{FWHM}} \gtrsim 0.23$). The most important difference from the case with Lorentzian coupling density is that the width of the split peaks decreases with the coupling strength, and for strong coupling it is dominated by the ensemble's homogeneous width or the cavity $\kappa$, as has been pointed out in Refs.~\cite{PhysRevA.53.2711, Grenoble}.

To estimate the position and width of the resonances, let us first consider the overdamped regime, $\Omega \ll \sigma$. The Taylor series expansion of Eq.~\eqref{eq:K+.gaussian} around $\xi=0$ reads
\begin{gather}
  \label{eq:K+.gauss.taylor}
  \tilde K^+(z) = \frac{\Omega^2}{\sqrt2 \sigma} \left[
    2\xi \sum_{n=0}^\infty \frac{(-2\xi^2)^n}{(2n+1)!!}
    -i\sqrt\pi \sum_{n=0}^\infty \frac{(-\xi^2)^n}{n!} \right].
\end{gather}
A good approximation for the first peak, which corresponds to the dressed cavity mode, can be readily obtained by keeping only the first term in both sums,
\begin{gather}
  \Re(E_+ - \omega_a) = \detuning /({1-\Omega^2/\sigma^2}),
  \\
  -2\Im E_+ = 2\kappa + \frac{2\kappa-\gamma_{\text{hom}}+\sqrt{2\pi}\sigma}
  {\sigma^2/\Omega^2-1}.
\end{gather}
For $\delta=\kappa=\gamma_{\text{hom}}=0$, (dash-dotted curve in Fig.~\ref{fig:peaks}d)
\begin{gather}
  \label{eq:ww.gauss}
  -2\Im E_+ = \sqrt{2\pi} \,
  {\Omega^2}/{\sigma} + \Ordo(\Omega^4)
\end{gather}
has quadratic dependence on the collective coupling~$\Omega$, which confirms the Weisskopf--Wigner approximation. Keeping the $\xi^2$ term in Eq.~\eqref{eq:K+.gauss.taylor}, we find a second root,
\begin{gather}
  \Re(E_- - \omega_a) = - \detuning + \Ordo (\Omega^2),
  \\
  -2\Im E_- = 2\gamma_{\text{hom}} - 2\kappa + \frac {8\sigma}{\sqrt{2\pi}} 
  \left( \frac{\sigma^2}{\Omega^2}-1 \right) + \Ordo (\Omega^2).
\end{gather}
It is worth mentioning that we have not only two, but infinitely many roots with increasingly large negative imaginary part (not shown in Fig.~\ref{fig:peaks}). This is a mathematical consequence of the dynamics of the subradiant modes and causes the cavity not to follow a simple exponential decay~\cite{WKM}. In the limit of weak coupling, however, the first root with finite imaginary part yields exponential decay in accord with the Weisskopf--Wigner approximation.

We consider now the strong coupling regime, $\Omega \gg \sigma$. The asymptotic expansion of Eq.~\eqref{eq:K+.gaussian} reads
\begin{gather}
  \label{eq:K+.gauss.asymptotic}
  \tilde K^+(z) \sim \frac{\Omega^2}{\sqrt2 \sigma} \left[
    -i\sqrt\pi e^{-\xi^2} + \frac1\xi \sum_{n=0}^\infty 
    \frac{(2n-1)!!}{(2\xi^2)^n} \right].
\end{gather}
Although the asymptotic power series does not converge, for any given $\xi$, we still get a good approximation if we truncate the series at $n\lesssim |\xi|^2$. In fact, the strong coupling approximation in Sec.~\ref{sec:oscil-vs-decay} corresponds to keeping only the $n=0$ term of the series, and it yields eigenenergies similar to Eq.~\eqref{eq:E.Rabi} with $\gamma = \gamma_{\text{hom}}$, which are shown as dashed lines in Fig.~\ref{fig:peaks}c. Focusing on the resonant case $\detuning=0$ and $\kappa = \gamma_{\text{hom}}=0$, we first neglect the imaginary part in Eq.~\eqref{eq:K+.gauss.asymptotic} and truncate the series after the $n=1$ term. This gives the position of the peaks in the next order (dotted curves in Fig.~\ref{fig:peaks}c),
\begin{gather}
  \label{eq:ReE.gauss.1}
  \Re (E_\pm^{(1)} - \omega_a) = \pm \sqrt{\Omega^2 + \sigma^2},
\end{gather}
or equivalently, $\xi^{(1)}_\pm = \pm \sqrt{(\Omega^2 +\sigma^2) /2\sigma^2}$. To obtain the imaginary part of the eigenenergies, we look for the solutions in the form $\xi = {\xi^{(1)}_\pm +\epsilon}$. We write $\xi^{(1)}_\pm$ in place of $\xi$ in the exponent in Eq.~\eqref{eq:K+.gauss.asymptotic}, and we use the facts that $1/\xi \approx {1/\xi^{(1)}_\pm - \epsilon / \xi^{(1)2}_\pm}$ and $1/\xi^3 \approx {1/\xi^{(1)3}_\pm - 3\epsilon / \xi^{(1)4}_\pm}$. With these approximations we have for the imaginary parts (dotted curve in Fig.~\ref{fig:peaks}d)
\begin{gather}
  \label{eq:M2ImE.gauss.2}
  -2\Im E^{(2)}_\pm = \sqrt{\frac{2\pi}e} e^{-\frac{\Omega^2}{2\sigma^2}}
  \left[\frac{\Omega^2-\sigma^2}{2\sigma} + \Ordo(\Omega^{-2})\right].
\end{gather}

\section{Protective energy gap in the strong coupling regime}
\label{sec:strong}

Ideally, when a spectrally narrow ensemble is strongly coupled to a cavity, excitations undergo coherent Rabi oscillation between the cavity and the superradiant spin wave mode at a collectively enhanced Rabi frequency $\Omega$,  without involving the subradiant modes. In reality, however, the inhomogeneity in the spin transition frequencies will gradually mix in the subradiant modes, thus resulting in decoherence of the cavity-superradiant subspace. The time scale for the latter process can vary, depending not only on the ensemble's inhomogeneous width, but also on the structure of the inhomogeneity. For a Lorentzian coupling density profile, e.g., it is always dominated by the inverse of the ensemble's inhomogeneous width, as we have seen in Sec.~\ref{sec:lorentzian}. For other distributions, like the Gaussian, this ``decoherence'' time can be significantly longer---an effect conventionally explained by a gapping mechanism~\cite{WKM, PhysRevLett.103.010502}.  In this Section, we investigate how the strong coupling to the cavity prevents the mixing of the superradiant and subradiant modes, thus leading to narrowing of the line width of the superradiant spin wave mode. For the cavity to have an effect, we will assume that the cavity is not too far from resonance ($\detuning \ll \Omega^2/\deltaomega$).

\subsection{Appearance of the energy gap}
\label{sec:rot-wave-approx}

In the absence of inhomogeneity, as shown in Sec.~\ref{sec:lorentzian}, the ``$+$'' and ``$-$'' eigenmodes are combinations of the superradiant spin wave mode and the cavity mode,
\begin{gather}
  \label{eq:phi0+}
  \hat\Phi^{(0)}_+  = \cos\tfrac\theta2\, \hat a_c + \sin\tfrac\theta2\, \hat b,\\
  \label{eq:phi0-}
  \hat\Phi^{(0)}_- = \sin\tfrac\theta2\, \hat a_c - \cos\tfrac\theta2\, \hat b,
\end{gather}
with the eigenenergies $E^{(0)}_\pm = (\omegabar + \omega_c)/2 \pm \Omega_R$. In the presence of inhomogeneity, however, these polariton modes are only approximate eigenmodes.  Their equations of motion read
\begin{gather}
  \label{eq:Phi+(0).rwa}
  \frac d{dt} \hat\Phi^{(0)}_+ = -i E^{(0)}_+ \hat\Phi^{(0)}_+
  -i \deltaomega \sin\tfrac\theta2\, \hat c
  , \\
  \label{eq:Phi-(0).rwa}
  \frac d{dt} \hat\Phi^{(0)}_- = -i E^{(0)}_- \hat\Phi^{(0)}_-
  +i \deltaomega \cos\tfrac\theta2 \, \hat c
  ,
\end{gather}
where we omitted the Langevin noise terms, $\hat c \equiv \deltaomega^{-1} \sum_j (\omega_j-\omegabar) \alpha_j^* \hat a_j$ is an annihilation operator corresponding to a subradiant mode orthogonal to both $\hat a_c$ and $\hat b$, and the ensemble's inhomogeneous width $\deltaomega$ is defined as the variance of the spin transition frequencies,
\begin{gather}
  \label{eq:deltaomega-def}
  \deltaomega^2 \equiv \sum_j
  |\omega_j - \omegabar|^2 |\alpha_j|^2
  = \int \big(\omega-\Re\omegabar\big)^2
  \frac{\rho(\omega)}{\Omega^2} \,d\omega,\\
  \omegabar \equiv \sum_j \omega_j |\alpha_j|^2
  = \int \big( \omega -\tfrac i2 \gamma_{\text{hom}} \big)
  \frac{\rho(\omega)}{\Omega^2} \,d\omega,
\end{gather}
where we assumed $\Im\omega_j = \Im\omegabar = -\frac12 \gamma_{\text{hom}}$ for all spins.

\begin{figure}
  \centering
  \includegraphics[scale=.5]{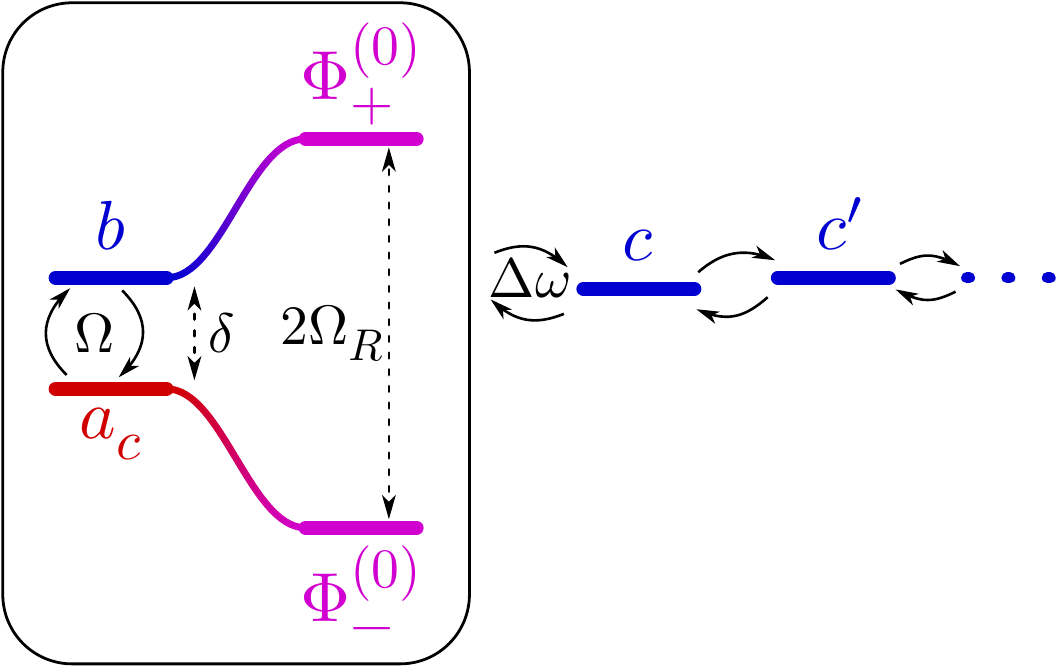}
  \caption{(Color online) 
    Level diagram showing the quasi-closed subspace consisting of the
    dressed polariton modes (superpositions of the cavity and the
    superradiant spin wave mode) coupled to a chain of subradiant spin
    wave modes. In the strong coupling regime $\Omega_R \gg
    \deltaomega$, the coupling between the dressed modes $\Phi_\pm$
    and the subradiant mode $c$ is off-resonant, therefore, the
    inhomogeneity does not induce decoherence of the dressed modes:
    the dressed levels are not broadened.
  }
  \label{fig:gapping}
\end{figure}

For a cavity not too far from resonance ($\detuning \ll \Omega^2/\deltaomega$), the limit of strong coupling is identified by the condition $\Omega \gg \deltaomega$. If the tail of the coupling density profile falls off sufficiently fast so that the Rabi-split eigenenergies $E^{(0)}_\pm$ lie far from all the spin transition frequencies, then the operators $\hat\Phi^{(0)}_\pm$ rotate fast with respect to $\hat c$, and the contribution of $\hat c$ in Eqs.~\eqref{eq:Phi+(0).rwa} and~\eqref{eq:Phi-(0).rwa} can be neglected (rotating wave approximation).  In other words, the spin dephasing processes induced by the inhomogeneous broadening have to bridge the energy gap between the dressed modes and the subradiant modes, and if this energy gap is large enough, the dressed modes are efficiently protected from decoherence (see Fig.~\ref{fig:gapping}).

Finally, we note that the gapping mechanism may be observed even when the cavity is out of resonance ($\deltaomega \ll \Omega \ll \detuning \ll \Omega^2 / \deltaomega$). Provided that the tail of the coupling density profile falls off sufficiently fast, the cavity can be adiabatically eliminated in the Born--Markov approximation. In this case, the cavity only has a dispersive effect, and its presence results in an energy shift of the superradiant mode equivalent to the ac Stark shift in optics.  Namely, $\hat H_1$ can be replaced by
\begin{gather}
  \label{eq:Hgap}
  \hat H_{\text{gap}} = \deltagap\, \hat b^\dag \hat b,
  \qquad \deltagap = -\frac{\Omega^2}{\detuning},
\end{gather}
independent of the state of the cavity. The superradiant mode, $\hat b^\dag \equiv \sum_j \alpha_j \hat a_j^\dag$, is an eigenmode of $\hat H_{\text{gap}}$, but due to the dephasing effects of the inhomogeneity in $\hat H_0$, it gets mixed with the subradiant modes, as shown by the Heisenberg equation of motion
\begin{gather}
  \frac d{dt} \hat b^\dag = i[\hat H_0 + \hat H_{\text{gap}}, \hat b^\dag]
  = i (\omegabar + \deltagap) \hat b^\dag + i \deltaomega \,\hat c^\dag.
\end{gather}
For a large gap ($\deltagap \gg \deltaomega$) and provided that the tail of the coupling density profile falls off sufficiently fast, the superradiant mode is energetically separated from the subradiant modes. Therefore, the inhomogeneous broadening cannot induce real transitions involving the superradiant mode. This energy gap may efficiently protect the quantum information stored in the superradiant spin wave mode from dephasing effects of the inhomogeneous broadening or spin diffusion~\cite{PhysRevLett.103.010502}.

\subsection{Corrections to the eigenenergies}
\label{sec:Epm-corr}

In this section, we estimate how fast the tail of the coupling density distribution should fall off in order for the cavity-superradiant subspace to become protected and for the line narrowing to manifest. In the strong coupling limit, the extremal eigenenergies $E_\pm$ lie far away from the typical spin transitions frequencies, and in this region ($|z - \omegabar| \gg \deltaomega$) we make an asymptotic power series expansion of $\tilde K^+(z)$ in the form
\begin{gather}
  \label{eq:K.series}
  \tilde K^+(z) =
  \sum_{k=0}^n \frac {\Omega^2 A_k}{(z - \omegabar)^{k+1}}
  + \Ordo \left(  \frac1{(z - \omegabar)^{n+2}} \right).
\end{gather}
For the Lorentzian \eqref{eq:Breit-Wigner}, $A_n = (-\tfrac i2 \gamma)^n$, while $A_{2n} = (2n-1)!!\sigma^{2n}$ and $A_{2n+1}=0$ for the Gaussian \eqref{eq:rho.gaussian}. In general, inserting the identity 
\begin{gather}
  \label{eq:reciproc-series}
  \mathcal P \frac1{\omega-\omega'} \sim \sum_{n=0}^\infty
  \frac{(\omega'-\Re\omegabar)^n}{(\omega-\Re\omegabar)^{n+1}}
\end{gather}
into \eqref{eq:Deltac} yields the asymptotic power series expansion
\begin{gather}
  \label{eq:Deltac.series}
  \Delta_c(\omega - \tfrac i2 \gamma_{\text{hom}}) \sim 
  \sum_{n=0}^\infty \frac {\Omega^2 M_n}{(\omega - \Re\omegabar)^{n+1}} ,
\end{gather}
as long as the statistical moments of the coupling density,
\begin{gather}
  \label{eq:Mn.moments}
  M_n \equiv \frac1{\Omega^{2}} \int ({\omega - \Re \omegabar})^n \rho(\omega) \, d\omega,
\end{gather}
are well-defined. In what follows, we will assume that the coupling density also has an asymptotic power series expansion,
\begin{gather}
  \label{eq:rho.series}
  \rho(\omega) \sim \sum_{n=1}^\infty \frac {\Omega^2 B_n}{(\omega - \Re \omegabar)^{n+1}}.
\end{gather}
Then $\Im A_k = -\pi B_k$. However, if the $n$-th moment \eqref{eq:Mn.moments} exists, we must have $B_k=0$, and $A_k=M_k$ for all $k\le n$.

Given the power series expansion \eqref{eq:K.series}, we now look for the resonance peaks in an iterative way. Since $M_0=1$, we start with $\tilde K^{+(0)}(z) \approx \Omega^2/({z - \omegabar})$ in the zeroth order. Equation \eqref{eq:resonance-peaks} then yields the same roots as those given by Eqs.~\eqref{eq:E.Rabi}--\eqref{eq:mixing-angle-complex} with $\gamma=\gamma_{\text{hom}}$, namely
\begin{gather}
  \label{eq:E0}
  E^{(0)}_\pm - \omegabar = \pm \Omega \big( \cot \tfrac\theta2 \big)^{\pm1}.
\end{gather}
In the first order, we look for the solution in the form $z = {E^{(0)} + \epsilon}$ and write
\begin{gather}
  \label{eq:K.series.1}
  \tilde K^{+(1)}(z) =
  \frac {\Omega^2}{E^{(0)} - \omegabar} \left[ 1
    +\frac{A_1-\epsilon}{E^{(0)} - \omegabar} \right].
\end{gather}
Then we obtain the corrections to $E^{(0)}_+$ and $E^{(0)}_-$,
\begin{gather}
  \label{eq:epsilon.1}
  \epsilon^{(1)}_+ = A_1 \sin^2\tfrac\theta2,
  \qquad
  \epsilon^{(1)}_- = A_1 \cos^2\tfrac\theta2.
\end{gather}

In the case of a Lorentzian coupling density, $A_1 = -\tfrac i2 \gamma$ immediately leads to wide resonance peaks dominated by the ensemble's inhomogeneous width $\gamma$. In general, if the tail of the coupling density profile falls off as $(\omega-\omegabar)^{-2}$, then the imaginary part of the correction $\epsilon^{(1)}$ reads
\begin{gather}
  -2\Im \epsilon^{(1)}_\pm = 2\pi B_1
  \left\{\begin{matrix}\sin^2\tfrac\theta2\\ \cos^2\tfrac\theta2 \end{matrix}\right\},
\end{gather}
and it is not reduced by a strong coupling constant~$\Omega$.  If, however, the first moment of $\rho(\omega)$ exists (and $\omegabar$ is such that $M_1=0$) and $\rho(\omega)$ has the asymptotic power series expansion \eqref{eq:rho.series}, then $B_1$ must be zero, and so $A_1=0$. In this case, we have to proceed further in the expansion.

In the second order, we assume that $A_1=0$, so we look for the solution in the form $z={E^{(0)}-\epsilon}$ and write
\begin{gather}
  \label{eq:K.series.2}
  \tilde K^{+(2)}(z) =
  \frac {\Omega^2}{E^{(0)} - \omegabar} \left[ 1
    -\frac\epsilon{E^{(0)} - \omegabar}
    +\frac{\epsilon^2+A_2}{(E^{(0)} - \omegabar)^2}
  \right].
\end{gather}
To simplify the roots of the resulting quadratic equation, we assume that $|A_2|\ll \Omega$ and make a series expansion with respect to $A_2/\Omega$. Then we obtain for the corrections
\begin{gather}
  \epsilon^{(2)}_\pm = \pm\frac{A_2^2}\Omega
  \left\{\begin{matrix}
      \sin^2\tfrac\theta2 \tan\tfrac\theta2\\
      \cos^2\tfrac\theta2 \cot\tfrac\theta2
    \end{matrix}\right\}
  + \Ordo (A_2^2/\Omega^4),
\end{gather}
and the resonance peaks are at
\begin{align}
  E_+^{(2)} &\approx \frac{\omegabar + \omega_c}2
  + \Omega_R \left[ 1 + \left(2\sin^4\tfrac\theta2 \right)
    \frac{A_2^2}{\Omega^2} \right],\\
  E_-^{(2)} &\approx \frac{\omegabar + \omega_c}2
  - \Omega_R \left[ 1 + \left(2\cos^4\tfrac\theta2 \right)
    \frac{A_2^2}{\Omega^2} \right].
\end{align}
We see that the contribution from $A_2$ is suppressed for large $\Omega$. If the second moment of $\rho(\omega)$ exists, then $A_2 = \deltaomega^2$ is pure real, and corrections to the resonance widths may only come from a complex mixing angle $\theta$. As an example, let us consider the resonant case $\detuning = 0$. The Rabi frequency is reduced, $\Omega_R = [{\Omega^2 - (\tfrac12 \gamma_{\text{hom}} - \kappa)^2 /4}]^{1/2}$, and the mixing angle is $\theta \approx \pi/2 - i(\tfrac12 \gamma_{\text{hom}} - \kappa) /2\Omega_R$. The real and imaginary parts of the two poles then read
\begin{gather}
  \Re(E_\pm^{(2)} - \omegabar) = \pm \Omega_R \left( 1
    + \frac12 \frac{\deltaomega^2}{\Omega^2} \right),\\
  \label{eq:M2ImE.gen.2}
  -2\Im E_\pm^{(2)} = \tfrac 12 \gamma_{\text{hom}} + \kappa
  +\big(\tfrac 12 \gamma_{\text{hom}} - \kappa \big)
   \frac{\deltaomega^2}{\Omega^2},
\end{gather}
A small inhomogeneous broadening enhances the Rabi splitting, while the width of the peaks is predominantly determined by the homogeneous width, and not the inhomogeneous one. We mention that in contrast to Eq.~\eqref{eq:M2ImE.gen.2}, Eq.~\eqref{eq:M2ImE.gauss.2} was obtained for $\kappa = \gamma_{\text{hom}} = 0$ and, therefore, we had to consider higher orders of $\rho(\omega)$ to estimate the peak widths.

\subsection{Losses due to inhomogeneous broadening}

To estimate the magnitude of the leakage from the two-dimensional subspace of the dressed states due to the inhomogeneous broadening, consider an initial excitation in the mode \eqref{eq:phi0+}.  The exact time evolution of this mode in the Heisenberg picture reads
\begin{gather}
  \hat\Phi^{(0)}_+ (t) = \sum_q \phi_q e^{-iE_qt} \hat\Phi_q,
\end{gather}
with $\phi_q = \cos\tfrac\theta2 \eta_{qc} + \sin\tfrac\theta2 \eta_{qs}$ and ${q=+}$,~$-$, $1$, \ldots, ${K-1}$.  We expect that the transition amplitude between the original and the evolved modes, given by
\begin{gather}
  \label{eq:G+(t)}
  G_{++}(t) \equiv
  \left\langle \big[\hat\Phi^{(0)}_+(t),
    \hat\Phi^{(0)\dag}_+(0)\big] \right\rangle
  = \sum_q |\phi_q|^2 e^{-iE_qt},
\end{gather}
rotates with an amplitude that is only slightly decreased compared to the homogeneous case.  Indeed, we can place a lower bound on $|G_{++}(t)|$ using the triangle inequality,
\begin{gather}
  \label{eq:|G+(t)|}
  |G_{++}(t)| \ge |\phi_+|^2 - \sum_{q\ne+} |\phi_q|^2 = 2 |\phi_+|^2 - 1.
\end{gather}
Using the method of Sec.~\ref{sec:Epm-corr}, we obtain
\begin{align}
  \eta_{+c} &\approx \cos \tfrac\theta2 \left[ 1 -
    \sin^4\tfrac\theta2 (2 + \sec^2\tfrac\theta2)
    \frac{\deltaomega^2}{2\Omega^2} \right],
  \displaybreak[0]\\
  \eta_{+s} &\approx \sin \tfrac\theta2 \left[ 1 +
    \tan^2\tfrac\theta2 \cos\theta (1+\cos^2\tfrac\theta2)
    \frac{\deltaomega^2}{2\Omega^2} \right],
\end{align}
and the inequality \eqref{eq:|G+(t)|} reads, to second order in $\deltaomega/\Omega$,
\begin{gather}
  |G_{++}(t)| \ge 2 |\phi_+|^2 - 1 \approx
  1 - 2 \Re \big(\sin^2\tfrac\theta2 \tan^2\tfrac\theta2 \big)
  \frac{\deltaomega^2}{\Omega^2},
\end{gather}
which yields $|G_{++}(t)| \gtrsim {1 - {\deltaomega^2}/{\Omega^2}}$ for the resonant case $\detuning=0$, and $|G_{++}(t)| \gtrsim {1 - 4 {\deltaomega^2}/{\deltagap^2}}$ for the off-resonant gapping regime ($\detuning \ll \Omega^2 / \deltaomega$, i.e., $\deltaomega \ll \deltagap$).

The leakage ($1-|G_{++}(t)|$) is thus bounded from above by a quantity proportional to  $\deltaomega^2 / \Omega^2$.  A similar bound can be derived for a $\hat\Phi^{(0)}_-$ excitation. The bound given by Eq.~\eqref{eq:|G+(t)|} is exact and valid for arbitrary coupling density. We emphasize, however, that these bounds are obtained by considering only the inhomogeneous distribution of the spin transition frequencies.  Other relaxation mechanisms of the individual spins, such as spontaneous decay, dephasing, or spin diffusion, may still be significant sources of losses.

\section{Conclusion}
\label{sec:conclusion}

We have studied highly polarized, inhomogeneously broad spin ensembles interacting with a single mode of a cavity. Using the resolvent formalism, we have shown how the transmission spectrum of the cavity depends on the coupling density profile and how the coupling density can be obtained from the transmission spectrum.

The strong superradiant coupling provides an energy gap for both the cavity and the superradiant spin wave modes from the subradiant modes, which may, in certain cases, efficiently protect the superradiant polariton and decrease its linewidth. We have investigated the criterion for the appearance of this gapping mechanism in the presence of inhomogeneity, and provided corrections to the conventionally used picture of a driven two-level system. We have also considered two specific inhomogeneous coupling density profiles: Lorentzian and Gaussian. We have shown that the gapping mechanism does not work for the former, and the polariton linewidth does not decrease with the collective coupling strength. For the latter, however, the gapping mechanism can efficiently reduce the polariton linewidth to a limit depending only on the cavity linewidth and the ensemble's homogeneous linewidth. In general, we have found that the ensemble's coupling density $\rho(\omega)$ should fall off as $\omega^{-3}$ or faster in order for the gapping mechanism to manifest itself and line narrowing to take place.


\bibliography{refs}

\begin{thebibliography}{34}%
\makeatletter
\providecommand \@ifxundefined [1]{%
 \@ifx{#1\undefined}
}%
\providecommand \@ifnum [1]{%
 \ifnum #1\expandafter \@firstoftwo
 \else \expandafter \@secondoftwo
 \fi
}%
\providecommand \@ifx [1]{%
 \ifx #1\expandafter \@firstoftwo
 \else \expandafter \@secondoftwo
 \fi
}%
\providecommand \natexlab [1]{#1}%
\providecommand \enquote  [1]{``#1''}%
\providecommand \bibnamefont  [1]{#1}%
\providecommand \bibfnamefont [1]{#1}%
\providecommand \citenamefont [1]{#1}%
\providecommand \href@noop [0]{\@secondoftwo}%
\providecommand \href [0]{\begingroup \@sanitize@url \@href}%
\providecommand \@href[1]{\@@startlink{#1}\@@href}%
\providecommand \@@href[1]{\endgroup#1\@@endlink}%
\providecommand \@sanitize@url [0]{\catcode `\\12\catcode `\$12\catcode
  `\&12\catcode `\#12\catcode `\^12\catcode `\_12\catcode `\%12\relax}%
\providecommand \@@startlink[1]{}%
\providecommand \@@endlink[0]{}%
\providecommand \url  [0]{\begingroup\@sanitize@url \@url }%
\providecommand \@url [1]{\endgroup\@href {#1}{\urlprefix }}%
\providecommand \urlprefix  [0]{URL }%
\providecommand \Eprint [0]{\href }%
\providecommand \doibase [0]{http://dx.doi.org/}%
\providecommand \selectlanguage [0]{\@gobble}%
\providecommand \bibinfo  [0]{\@secondoftwo}%
\providecommand \bibfield  [0]{\@secondoftwo}%
\providecommand \translation [1]{[#1]}%
\providecommand \BibitemOpen [0]{}%
\providecommand \bibitemStop [0]{}%
\providecommand \bibitemNoStop [0]{.\EOS\space}%
\providecommand \EOS [0]{\spacefactor3000\relax}%
\providecommand \BibitemShut  [1]{\csname bibitem#1\endcsname}%
\let\auto@bib@innerbib\@empty
\bibitem [{\citenamefont {Hammerer}\ \emph {et~al.}(2010)\citenamefont
  {Hammerer}, \citenamefont {S{\o}rensen},\ and\ \citenamefont
  {Polzik}}]{RevModPhys.82.1041}%
  \BibitemOpen
  \bibfield  {author} {\bibinfo {author} {\bibfnamefont {K.}~\bibnamefont
  {Hammerer}}, \bibinfo {author} {\bibfnamefont {A.~S.}\ \bibnamefont
  {S{\o}rensen}}, \ and\ \bibinfo {author} {\bibfnamefont {E.~S.}\ \bibnamefont
  {Polzik}},\ }\href {\doibase 10.1103/RevModPhys.82.1041} {\bibfield
  {journal} {\bibinfo  {journal} {Rev. Mod. Phys.}\ }\textbf {\bibinfo {volume}
  {82}},\ \bibinfo {pages} {1041} (\bibinfo {year} {2010})}\BibitemShut
  {NoStop}%
\bibitem [{\citenamefont {Fleischhauer}\ and\ \citenamefont
  {Lukin}(2000)}]{PhysRevLett.84.5094}%
  \BibitemOpen
  \bibfield  {author} {\bibinfo {author} {\bibfnamefont {M.}~\bibnamefont
  {Fleischhauer}}\ and\ \bibinfo {author} {\bibfnamefont {M.~D.}\ \bibnamefont
  {Lukin}},\ }\href {\doibase 10.1103/PhysRevLett.84.5094} {\bibfield
  {journal} {\bibinfo  {journal} {Phys. Rev. Lett.}\ }\textbf {\bibinfo
  {volume} {84}},\ \bibinfo {pages} {5094} (\bibinfo {year}
  {2000})}\BibitemShut {NoStop}%
\bibitem [{\citenamefont {Duan}\ \emph {et~al.}(2001)\citenamefont {Duan},
  \citenamefont {Lukin}, \citenamefont {Cirac},\ and\ \citenamefont
  {Zoller}}]{Nature.414.413}%
  \BibitemOpen
  \bibfield  {author} {\bibinfo {author} {\bibfnamefont {L.-M.}\ \bibnamefont
  {Duan}}, \bibinfo {author} {\bibfnamefont {M.~D.}\ \bibnamefont {Lukin}},
  \bibinfo {author} {\bibfnamefont {J.~I.}\ \bibnamefont {Cirac}}, \ and\
  \bibinfo {author} {\bibfnamefont {P.}~\bibnamefont {Zoller}},\ }\href
  {\doibase 10.1038/35106500} {\bibfield  {journal} {\bibinfo  {journal}
  {Nature}\ }\textbf {\bibinfo {volume} {414}},\ \bibinfo {pages} {413}
  (\bibinfo {year} {2001})}\BibitemShut {NoStop}%
\bibitem [{\citenamefont {Carlson}\ \emph {et~al.}(1983)\citenamefont
  {Carlson}, \citenamefont {Rothberg}, \citenamefont {Yodh}, \citenamefont
  {Babbitt},\ and\ \citenamefont {Mossberg}}]{OptLett.8.483}%
  \BibitemOpen
  \bibfield  {author} {\bibinfo {author} {\bibfnamefont {N.~W.}\ \bibnamefont
  {Carlson}}, \bibinfo {author} {\bibfnamefont {L.~J.}\ \bibnamefont
  {Rothberg}}, \bibinfo {author} {\bibfnamefont {A.~G.}\ \bibnamefont {Yodh}},
  \bibinfo {author} {\bibfnamefont {W.~R.}\ \bibnamefont {Babbitt}}, \ and\
  \bibinfo {author} {\bibfnamefont {T.~W.}\ \bibnamefont {Mossberg}},\ }\href
  {\doibase 10.1364/OL.8.000483} {\bibfield  {journal} {\bibinfo  {journal}
  {Opt. Lett.}\ }\textbf {\bibinfo {volume} {8}},\ \bibinfo {pages} {483}
  (\bibinfo {year} {1983})}\BibitemShut {NoStop}%
\bibitem [{\citenamefont {Nilsson}\ and\ \citenamefont
  {Kr{\"o}ll}(2005)}]{OptCommun.247.393}%
  \BibitemOpen
  \bibfield  {author} {\bibinfo {author} {\bibfnamefont {M.}~\bibnamefont
  {Nilsson}}\ and\ \bibinfo {author} {\bibfnamefont {S.}~\bibnamefont
  {Kr{\"o}ll}},\ }\href {\doibase 10.1016/j.optcom.2004.11.077} {\bibfield
  {journal} {\bibinfo  {journal} {Opt. Commun.}\ }\textbf {\bibinfo {volume}
  {247}},\ \bibinfo {pages} {393} (\bibinfo {year} {2005})}\BibitemShut
  {NoStop}%
\bibitem [{\citenamefont {Kraus}\ \emph {et~al.}(2006)\citenamefont {Kraus},
  \citenamefont {Tittel}, \citenamefont {Gisin}, \citenamefont {Nilsson},
  \citenamefont {Kroll},\ and\ \citenamefont {Cirac}}]{PhysRevA.73.020302}%
  \BibitemOpen
  \bibfield  {author} {\bibinfo {author} {\bibfnamefont {B.}~\bibnamefont
  {Kraus}}, \bibinfo {author} {\bibfnamefont {W.}~\bibnamefont {Tittel}},
  \bibinfo {author} {\bibfnamefont {N.}~\bibnamefont {Gisin}}, \bibinfo
  {author} {\bibfnamefont {M.}~\bibnamefont {Nilsson}}, \bibinfo {author}
  {\bibfnamefont {S.}~\bibnamefont {Kroll}}, \ and\ \bibinfo {author}
  {\bibfnamefont {J.~I.}\ \bibnamefont {Cirac}},\ }\href {\doibase
  10.1103/PhysRevA.73.020302} {\bibfield  {journal} {\bibinfo  {journal} {Phys.
  Rev. A}\ }\textbf {\bibinfo {volume} {73}},\ \bibinfo {pages} {020302(R)}
  (\bibinfo {year} {2006})}\BibitemShut {NoStop}%
\bibitem [{\citenamefont {Afzelius}\ \emph {et~al.}(2009)\citenamefont
  {Afzelius}, \citenamefont {Simon}, \citenamefont {de~Riedmatten},\ and\
  \citenamefont {Gisin}}]{PhysRevA.79.052329}%
  \BibitemOpen
  \bibfield  {author} {\bibinfo {author} {\bibfnamefont {M.}~\bibnamefont
  {Afzelius}}, \bibinfo {author} {\bibfnamefont {C.}~\bibnamefont {Simon}},
  \bibinfo {author} {\bibfnamefont {H.}~\bibnamefont {de~Riedmatten}}, \ and\
  \bibinfo {author} {\bibfnamefont {N.}~\bibnamefont {Gisin}},\ }\href
  {\doibase 10.1103/PhysRevA.79.052329} {\bibfield  {journal} {\bibinfo
  {journal} {Phys. Rev. A}\ }\textbf {\bibinfo {volume} {79}},\ \bibinfo
  {pages} {052329} (\bibinfo {year} {2009})}\BibitemShut {NoStop}%
\bibitem [{\citenamefont {Afzelius}\ \emph {et~al.}(2010)\citenamefont
  {Afzelius}, \citenamefont {Usmani}, \citenamefont {Amari}, \citenamefont
  {Lauritzen}, \citenamefont {Walther}, \citenamefont {Simon}, \citenamefont
  {Sangouard}, \citenamefont {Min{\'{a}}{\v{r}}}, \citenamefont
  {de~Riedmatten}, \citenamefont {Gisin},\ and\ \citenamefont
  {Kr{\"{o}}ll}}]{PhysRevLett.104.040503}%
  \BibitemOpen
  \bibfield  {author} {\bibinfo {author} {\bibfnamefont {M.}~\bibnamefont
  {Afzelius}}, \bibinfo {author} {\bibfnamefont {I.}~\bibnamefont {Usmani}},
  \bibinfo {author} {\bibfnamefont {A.}~\bibnamefont {Amari}}, \bibinfo
  {author} {\bibfnamefont {B.}~\bibnamefont {Lauritzen}}, \bibinfo {author}
  {\bibfnamefont {A.}~\bibnamefont {Walther}}, \bibinfo {author} {\bibfnamefont
  {C.}~\bibnamefont {Simon}}, \bibinfo {author} {\bibfnamefont
  {N.}~\bibnamefont {Sangouard}}, \bibinfo {author} {\bibfnamefont
  {J.}~\bibnamefont {Min{\'{a}}{\v{r}}}}, \bibinfo {author} {\bibfnamefont
  {H.}~\bibnamefont {de~Riedmatten}}, \bibinfo {author} {\bibfnamefont
  {N.}~\bibnamefont {Gisin}}, \ and\ \bibinfo {author} {\bibfnamefont
  {S.}~\bibnamefont {Kr{\"{o}}ll}},\ }\href {\doibase
  10.1103/PhysRevLett.104.040503} {\bibfield  {journal} {\bibinfo  {journal}
  {Phys. Rev. Lett.}\ }\textbf {\bibinfo {volume} {104}},\ \bibinfo {pages}
  {040503} (\bibinfo {year} {2010})}\BibitemShut {NoStop}%
\bibitem [{\citenamefont {Bonarota}\ and\ \citenamefont {J.-L.
  Le~Gou{\"e}t}()}]{arXiv.1009.2317}%
  \BibitemOpen
  \bibfield  {author} {\bibinfo {author} {\bibfnamefont {M.}~\bibnamefont
  {Bonarota}}\ and\ \bibinfo {author} {\bibfnamefont {T.~C.}\ \bibnamefont
  {J.-L. Le~Gou{\"e}t}},\ }\href@noop {} {\enquote {\bibinfo {title} {Highly
  multimode memory in a crystal},}\ }\bibinfo {note}
  {ArXiv:1009.2317v2}\BibitemShut {NoStop}%
\bibitem [{\citenamefont {Brennecke}\ \emph {et~al.}(2007)\citenamefont
  {Brennecke}, \citenamefont {Donner}, \citenamefont {Ritter}, \citenamefont
  {Bourdel}, \citenamefont {K{\"o}hl},\ and\ \citenamefont
  {Esslinger}}]{Nature.450.268}%
  \BibitemOpen
  \bibfield  {author} {\bibinfo {author} {\bibfnamefont {F.}~\bibnamefont
  {Brennecke}}, \bibinfo {author} {\bibfnamefont {T.}~\bibnamefont {Donner}},
  \bibinfo {author} {\bibfnamefont {S.}~\bibnamefont {Ritter}}, \bibinfo
  {author} {\bibfnamefont {T.}~\bibnamefont {Bourdel}}, \bibinfo {author}
  {\bibfnamefont {M.}~\bibnamefont {K{\"o}hl}}, \ and\ \bibinfo {author}
  {\bibfnamefont {T.}~\bibnamefont {Esslinger}},\ }\href {\doibase
  10.1038/nature06120} {\bibfield  {journal} {\bibinfo  {journal} {Nature}\
  }\textbf {\bibinfo {volume} {450}},\ \bibinfo {pages} {268} (\bibinfo {year}
  {2007})}\BibitemShut {NoStop}%
\bibitem [{\citenamefont {Colombe}\ \emph {et~al.}(2007)\citenamefont
  {Colombe}, \citenamefont {Steinmetz}, \citenamefont {Dubois}, \citenamefont
  {Linke}, \citenamefont {Hunger},\ and\ \citenamefont
  {Reichel}}]{Nature.450.272}%
  \BibitemOpen
  \bibfield  {author} {\bibinfo {author} {\bibfnamefont {Y.}~\bibnamefont
  {Colombe}}, \bibinfo {author} {\bibfnamefont {T.}~\bibnamefont {Steinmetz}},
  \bibinfo {author} {\bibfnamefont {G.}~\bibnamefont {Dubois}}, \bibinfo
  {author} {\bibfnamefont {F.}~\bibnamefont {Linke}}, \bibinfo {author}
  {\bibfnamefont {D.}~\bibnamefont {Hunger}}, \ and\ \bibinfo {author}
  {\bibfnamefont {J.}~\bibnamefont {Reichel}},\ }\href {\doibase
  10.1038/nature06331} {\bibfield  {journal} {\bibinfo  {journal} {Nature}\
  }\textbf {\bibinfo {volume} {450}},\ \bibinfo {pages} {272} (\bibinfo {year}
  {2007})}\BibitemShut {NoStop}%
\bibitem [{\citenamefont {Herskind}\ \emph {et~al.}(2009)\citenamefont
  {Herskind}, \citenamefont {Dantan}, \citenamefont {Marler}, \citenamefont
  {Albert},\ and\ \citenamefont {Drewsen}}]{NatPhys.5.494}%
  \BibitemOpen
  \bibfield  {author} {\bibinfo {author} {\bibfnamefont {P.~F.}\ \bibnamefont
  {Herskind}}, \bibinfo {author} {\bibfnamefont {A.}~\bibnamefont {Dantan}},
  \bibinfo {author} {\bibfnamefont {J.~P.}\ \bibnamefont {Marler}}, \bibinfo
  {author} {\bibfnamefont {M.}~\bibnamefont {Albert}}, \ and\ \bibinfo {author}
  {\bibfnamefont {M.}~\bibnamefont {Drewsen}},\ }\href {\doibase
  10.1038/NPHYS1302} {\bibfield  {journal} {\bibinfo  {journal} {Nat. Phys.}\
  }\textbf {\bibinfo {volume} {5}},\ \bibinfo {pages} {494} (\bibinfo {year}
  {2009})}\BibitemShut {NoStop}%
\bibitem [{\citenamefont {Andr{\'e}}\ \emph {et~al.}(2006)\citenamefont
  {Andr{\'e}}, \citenamefont {DeMille}, \citenamefont {Doyle}, \citenamefont
  {Lukin}, \citenamefont {Maxwell}, \citenamefont {Rabl}, \citenamefont
  {Schoelkopf},\ and\ \citenamefont {Zoller}}]{NatPhys.2.636}%
  \BibitemOpen
  \bibfield  {author} {\bibinfo {author} {\bibfnamefont {A.}~\bibnamefont
  {Andr{\'e}}}, \bibinfo {author} {\bibfnamefont {D.}~\bibnamefont {DeMille}},
  \bibinfo {author} {\bibfnamefont {J.~M.}\ \bibnamefont {Doyle}}, \bibinfo
  {author} {\bibfnamefont {M.~D.}\ \bibnamefont {Lukin}}, \bibinfo {author}
  {\bibfnamefont {S.~E.}\ \bibnamefont {Maxwell}}, \bibinfo {author}
  {\bibfnamefont {P.}~\bibnamefont {Rabl}}, \bibinfo {author} {\bibfnamefont
  {R.~J.}\ \bibnamefont {Schoelkopf}}, \ and\ \bibinfo {author} {\bibfnamefont
  {P.}~\bibnamefont {Zoller}},\ }\href {\doibase doi:10.1038/nphys386}
  {\bibfield  {journal} {\bibinfo  {journal} {Nat. Phys.}\ }\textbf {\bibinfo
  {volume} {2}},\ \bibinfo {pages} {636} (\bibinfo {year} {2006})}\BibitemShut
  {NoStop}%
\bibitem [{\citenamefont {Tordrup}\ \emph {et~al.}(2008)\citenamefont
  {Tordrup}, \citenamefont {Negretti},\ and\ \citenamefont
  {M\o{}lmer}}]{PhysRevLett.101.040501}%
  \BibitemOpen
  \bibfield  {author} {\bibinfo {author} {\bibfnamefont {K.}~\bibnamefont
  {Tordrup}}, \bibinfo {author} {\bibfnamefont {A.}~\bibnamefont {Negretti}}, \
  and\ \bibinfo {author} {\bibfnamefont {K.}~\bibnamefont {M\o{}lmer}},\ }\href
  {\doibase 10.1103/PhysRevLett.101.040501} {\bibfield  {journal} {\bibinfo
  {journal} {Phys. Rev. Lett.}\ }\textbf {\bibinfo {volume} {101}},\ \bibinfo
  {pages} {040501} (\bibinfo {year} {2008})}\BibitemShut {NoStop}%
\bibitem [{\citenamefont {Wesenberg}\ \emph {et~al.}(2009)\citenamefont
  {Wesenberg}, \citenamefont {Ardavan}, \citenamefont {Briggs}, \citenamefont
  {Morton}, \citenamefont {Schoelkopf}, \citenamefont {Schuster},\ and\
  \citenamefont {M{\o}lmer}}]{PhysRevLett.103.070502}%
  \BibitemOpen
  \bibfield  {author} {\bibinfo {author} {\bibfnamefont {J.~H.}\ \bibnamefont
  {Wesenberg}}, \bibinfo {author} {\bibfnamefont {A.}~\bibnamefont {Ardavan}},
  \bibinfo {author} {\bibfnamefont {G.~A.~D.}\ \bibnamefont {Briggs}}, \bibinfo
  {author} {\bibfnamefont {J.~J.~L.}\ \bibnamefont {Morton}}, \bibinfo {author}
  {\bibfnamefont {R.~J.}\ \bibnamefont {Schoelkopf}}, \bibinfo {author}
  {\bibfnamefont {D.~I.}\ \bibnamefont {Schuster}}, \ and\ \bibinfo {author}
  {\bibfnamefont {K.}~\bibnamefont {M{\o}lmer}},\ }\href {\doibase
  10.1103/PhysRevLett.103.070502} {\bibfield  {journal} {\bibinfo  {journal}
  {Phys. Rev. Lett.}\ }\textbf {\bibinfo {volume} {103}},\ \bibinfo {pages}
  {070502} (\bibinfo {year} {2009})}\BibitemShut {NoStop}%
\bibitem [{\citenamefont {Schuster}\ \emph {et~al.}(2010)\citenamefont
  {Schuster}, \citenamefont {Sears}, \citenamefont {Ginossar}, \citenamefont
  {Dicarlo}, \citenamefont {Frunzio}, \citenamefont {Morton}, \citenamefont
  {Wu}, \citenamefont {Briggs}, \citenamefont {Buckley}, \citenamefont
  {Awschalom},\ and\ \citenamefont {Schoelkopf}}]{PhysRevLett.105.140501}%
  \BibitemOpen
  \bibfield  {author} {\bibinfo {author} {\bibfnamefont {D.~I.}\ \bibnamefont
  {Schuster}}, \bibinfo {author} {\bibfnamefont {A.~P.}\ \bibnamefont {Sears}},
  \bibinfo {author} {\bibfnamefont {E.}~\bibnamefont {Ginossar}}, \bibinfo
  {author} {\bibfnamefont {L.}~\bibnamefont {Dicarlo}}, \bibinfo {author}
  {\bibfnamefont {L.}~\bibnamefont {Frunzio}}, \bibinfo {author} {\bibfnamefont
  {J.~J.~L.}\ \bibnamefont {Morton}}, \bibinfo {author} {\bibfnamefont
  {H.}~\bibnamefont {Wu}}, \bibinfo {author} {\bibfnamefont {G.~A.~D.}\
  \bibnamefont {Briggs}}, \bibinfo {author} {\bibfnamefont {B.~B.}\
  \bibnamefont {Buckley}}, \bibinfo {author} {\bibfnamefont {D.~D.}\
  \bibnamefont {Awschalom}}, \ and\ \bibinfo {author} {\bibfnamefont {R.~J.}\
  \bibnamefont {Schoelkopf}},\ }\href {\doibase 10.1103/PhysRevLett.105.140501}
  {\bibfield  {journal} {\bibinfo  {journal} {Phys. Rev. Lett.}\ }\textbf
  {\bibinfo {volume} {105}},\ \bibinfo {pages} {140501} (\bibinfo {year}
  {2010})}\BibitemShut {NoStop}%
\bibitem [{\citenamefont {Kubo}\ \emph {et~al.}(2010)\citenamefont {Kubo},
  \citenamefont {Ong}, \citenamefont {Bertet}, \citenamefont {Vion},
  \citenamefont {Jacques}, \citenamefont {Zheng}, \citenamefont {Dr{\'e}au},
  \citenamefont {Roch}, \citenamefont {Auffeves}, \citenamefont {Jelezko},
  \citenamefont {Wrachtrup}, \citenamefont {Barthe}, \citenamefont {Bergonzo},\
  and\ \citenamefont {Esteve}}]{PhysRevLett.105.140502}%
  \BibitemOpen
  \bibfield  {author} {\bibinfo {author} {\bibfnamefont {Y.}~\bibnamefont
  {Kubo}}, \bibinfo {author} {\bibfnamefont {F.~R.}\ \bibnamefont {Ong}},
  \bibinfo {author} {\bibfnamefont {P.}~\bibnamefont {Bertet}}, \bibinfo
  {author} {\bibfnamefont {D.}~\bibnamefont {Vion}}, \bibinfo {author}
  {\bibfnamefont {V.}~\bibnamefont {Jacques}}, \bibinfo {author} {\bibfnamefont
  {D.}~\bibnamefont {Zheng}}, \bibinfo {author} {\bibfnamefont
  {A.}~\bibnamefont {Dr{\'e}au}}, \bibinfo {author} {\bibfnamefont {J.~F.}\
  \bibnamefont {Roch}}, \bibinfo {author} {\bibfnamefont {A.}~\bibnamefont
  {Auffeves}}, \bibinfo {author} {\bibfnamefont {F.}~\bibnamefont {Jelezko}},
  \bibinfo {author} {\bibfnamefont {J.}~\bibnamefont {Wrachtrup}}, \bibinfo
  {author} {\bibfnamefont {M.~F.}\ \bibnamefont {Barthe}}, \bibinfo {author}
  {\bibfnamefont {P.}~\bibnamefont {Bergonzo}}, \ and\ \bibinfo {author}
  {\bibfnamefont {D.}~\bibnamefont {Esteve}},\ }\href {\doibase
  10.1103/PhysRevLett.105.140502} {\bibfield  {journal} {\bibinfo  {journal}
  {Phys. Rev. Lett.}\ }\textbf {\bibinfo {volume} {105}},\ \bibinfo {pages}
  {140502} (\bibinfo {year} {2010})}\BibitemShut {NoStop}%
\bibitem [{\citenamefont {R{\"o}hlsberger}\ \emph {et~al.}(2010)\citenamefont
  {R{\"o}hlsberger}, \citenamefont {Schlage}, \citenamefont {Sahoo},
  \citenamefont {Couet},\ and\ \citenamefont {R{\"u}ffer}}]{Science.328.1248}%
  \BibitemOpen
  \bibfield  {author} {\bibinfo {author} {\bibfnamefont {R.}~\bibnamefont
  {R{\"o}hlsberger}}, \bibinfo {author} {\bibfnamefont {K.}~\bibnamefont
  {Schlage}}, \bibinfo {author} {\bibfnamefont {B.}~\bibnamefont {Sahoo}},
  \bibinfo {author} {\bibfnamefont {S.}~\bibnamefont {Couet}}, \ and\ \bibinfo
  {author} {\bibfnamefont {R.}~\bibnamefont {R{\"u}ffer}},\ }\href {\doibase
  10.1126/science.1187770} {\bibfield  {journal} {\bibinfo  {journal}
  {Science}\ }\textbf {\bibinfo {volume} {328}},\ \bibinfo {pages} {1248}
  (\bibinfo {year} {2010})}\BibitemShut {NoStop}%
\bibitem [{\citenamefont {Wu}\ \emph {et~al.}(2010)\citenamefont {Wu},
  \citenamefont {George}, \citenamefont {Wesenberg}, \citenamefont {M{\o}lmer},
  \citenamefont {Schuster}, \citenamefont {Schoelkopf}, \citenamefont {Itoh},
  \citenamefont {Ardavan}, \citenamefont {Morton},\ and\ \citenamefont
  {Briggs}}]{PhysRevLett.105.140503}%
  \BibitemOpen
  \bibfield  {author} {\bibinfo {author} {\bibfnamefont {H.}~\bibnamefont
  {Wu}}, \bibinfo {author} {\bibfnamefont {R.~E.}\ \bibnamefont {George}},
  \bibinfo {author} {\bibfnamefont {J.~H.}\ \bibnamefont {Wesenberg}}, \bibinfo
  {author} {\bibfnamefont {K.}~\bibnamefont {M{\o}lmer}}, \bibinfo {author}
  {\bibfnamefont {D.~I.}\ \bibnamefont {Schuster}}, \bibinfo {author}
  {\bibfnamefont {R.~J.}\ \bibnamefont {Schoelkopf}}, \bibinfo {author}
  {\bibfnamefont {K.~M.}\ \bibnamefont {Itoh}}, \bibinfo {author}
  {\bibfnamefont {A.}~\bibnamefont {Ardavan}}, \bibinfo {author} {\bibfnamefont
  {J.~J.~L.}\ \bibnamefont {Morton}}, \ and\ \bibinfo {author} {\bibfnamefont
  {G.~A.~D.}\ \bibnamefont {Briggs}},\ }\href {\doibase
  10.1103/PhysRevLett.105.140503} {\bibfield  {journal} {\bibinfo  {journal}
  {Phys. Rev. Lett.}\ }\textbf {\bibinfo {volume} {105}},\ \bibinfo {pages}
  {140503} (\bibinfo {year} {2010})}\BibitemShut {NoStop}%
\bibitem [{\citenamefont {Breuer}\ \emph {et~al.}(2004)\citenamefont {Breuer},
  \citenamefont {Burgarth},\ and\ \citenamefont
  {Petruccione}}]{PhysRevB.70.045323}%
  \BibitemOpen
  \bibfield  {author} {\bibinfo {author} {\bibfnamefont {H.-P.}\ \bibnamefont
  {Breuer}}, \bibinfo {author} {\bibfnamefont {D.}~\bibnamefont {Burgarth}}, \
  and\ \bibinfo {author} {\bibfnamefont {F.}~\bibnamefont {Petruccione}},\
  }\href {\doibase 10.1103/PhysRevB.70.045323} {\bibfield  {journal} {\bibinfo
  {journal} {Phys. Rev. B}\ }\textbf {\bibinfo {volume} {70}},\ \bibinfo
  {pages} {045323} (\bibinfo {year} {2004})}\BibitemShut {NoStop}%
\bibitem [{\citenamefont {Cohen-Tannoudji}\ \emph {et~al.}(1992)\citenamefont
  {Cohen-Tannoudji}, \citenamefont {Dupont-Roc},\ and\ \citenamefont
  {Grynberg}}]{cct-api}%
  \BibitemOpen
  \bibfield  {author} {\bibinfo {author} {\bibfnamefont {C.}~\bibnamefont
  {Cohen-Tannoudji}}, \bibinfo {author} {\bibfnamefont {J.}~\bibnamefont
  {Dupont-Roc}}, \ and\ \bibinfo {author} {\bibfnamefont {G.}~\bibnamefont
  {Grynberg}},\ }\href@noop {} {\emph {\bibinfo {title} {Atom-photon
  interactions: basic processes and applications}}}\ (\bibinfo  {publisher}
  {John Wiley \& Sons},\ \bibinfo {year} {1992})\BibitemShut {NoStop}%
\bibitem [{\citenamefont {Peskin}\ and\ \citenamefont
  {Schroeder}(1995)}]{peskin-qft}%
  \BibitemOpen
  \bibfield  {author} {\bibinfo {author} {\bibfnamefont {M.~E.}\ \bibnamefont
  {Peskin}}\ and\ \bibinfo {author} {\bibfnamefont {D.~V.}\ \bibnamefont
  {Schroeder}},\ }\href@noop {} {\emph {\bibinfo {title} {Introduction to
  quantum field theory}}}\ (\bibinfo  {publisher} {Westview Press},\ \bibinfo
  {address} {Boulder, Colorado},\ \bibinfo {year} {1995})\BibitemShut {NoStop}%
\bibitem [{\citenamefont {Lee}(1954)}]{PhysRev.95.1329}%
  \BibitemOpen
  \bibfield  {author} {\bibinfo {author} {\bibfnamefont {T.~D.}\ \bibnamefont
  {Lee}},\ }\href {\doibase 10.1103/PhysRev.95.1329} {\bibfield  {journal}
  {\bibinfo  {journal} {Phys. Rev.}\ }\textbf {\bibinfo {volume} {95}},\
  \bibinfo {pages} {1329} (\bibinfo {year} {1954})}\BibitemShut {NoStop}%
\bibitem [{\citenamefont {Friedrichs}(1948)}]{CommunPureApplMath.1.361}%
  \BibitemOpen
  \bibfield  {author} {\bibinfo {author} {\bibfnamefont {K.~O.}\ \bibnamefont
  {Friedrichs}},\ }\href {\doibase 10.1002/cpa.3160010404} {\bibfield
  {journal} {\bibinfo  {journal} {Commun. Pure Appl. Math.}\ }\textbf {\bibinfo
  {volume} {1}},\ \bibinfo {pages} {361} (\bibinfo {year} {1948})}\BibitemShut
  {NoStop}%
\bibitem [{\citenamefont {Caianiello}(1961)}]{caianiello61}%
  \BibitemOpen
  \bibinfo {editor} {\bibfnamefont {E.~R.}\ \bibnamefont {Caianiello}},\ ed.,\
  \href@noop {} {\emph {\bibinfo {title} {Lectures on field theory and the
  many-body problem}}}\ (\bibinfo  {publisher} {Academic Press},\ \bibinfo
  {address} {New York},\ \bibinfo {year} {1961})\BibitemShut {NoStop}%
\bibitem [{\citenamefont {Stenholm}(2000)}]{OptCommun.179.247}%
  \BibitemOpen
  \bibfield  {author} {\bibinfo {author} {\bibfnamefont {S.}~\bibnamefont
  {Stenholm}},\ }\href {\doibase 10.1016/S0030-4018(99)00543-X} {\bibfield
  {journal} {\bibinfo  {journal} {Opt. Commun.}\ }\textbf {\bibinfo {volume}
  {179}},\ \bibinfo {pages} {247} (\bibinfo {year} {2000})}\BibitemShut
  {NoStop}%
\bibitem [{\citenamefont {Houdr{\'{e}}}\ \emph {et~al.}(1996)\citenamefont
  {Houdr{\'{e}}}, \citenamefont {Stanley},\ and\ \citenamefont
  {Ilegems}}]{PhysRevA.53.2711}%
  \BibitemOpen
  \bibfield  {author} {\bibinfo {author} {\bibfnamefont {R.}~\bibnamefont
  {Houdr{\'{e}}}}, \bibinfo {author} {\bibfnamefont {R.~P.}\ \bibnamefont
  {Stanley}}, \ and\ \bibinfo {author} {\bibfnamefont {M.}~\bibnamefont
  {Ilegems}},\ }\href {\doibase 10.1103/PhysRevA.53.2711} {\bibfield  {journal}
  {\bibinfo  {journal} {Phys. Rev. A}\ }\textbf {\bibinfo {volume} {53}},\
  \bibinfo {pages} {2711} (\bibinfo {year} {1996})}\BibitemShut {NoStop}%
\bibitem [{\citenamefont {Diniz}\ \emph {et~al.}()\citenamefont {Diniz},
  \citenamefont {Portolan}, \citenamefont {Ferreira}, \citenamefont
  {G{\'e}rard}, \citenamefont {Bertet},\ and\ \citenamefont
  {Auff{\`e}ves}}]{Grenoble}%
  \BibitemOpen
  \bibfield  {author} {\bibinfo {author} {\bibfnamefont {I.}~\bibnamefont
  {Diniz}}, \bibinfo {author} {\bibfnamefont {S.}~\bibnamefont {Portolan}},
  \bibinfo {author} {\bibfnamefont {R.}~\bibnamefont {Ferreira}}, \bibinfo
  {author} {\bibfnamefont {J.}~\bibnamefont {G{\'e}rard}}, \bibinfo {author}
  {\bibfnamefont {P.}~\bibnamefont {Bertet}}, \ and\ \bibinfo {author}
  {\bibfnamefont {A.}~\bibnamefont {Auff{\`e}ves}},\ }\href@noop {} {\enquote
  {\bibinfo {title} {Strongly coupling a cavity to inhomogeneous ensembles of
  emitters: potential for long lived solid-state quantum memories},}\ }\Eprint
  {http://arxiv.org/abs/arXiv:1101.1842v1} {arXiv:1101.1842v1} \BibitemShut
  {NoStop}%
\bibitem [{\citenamefont {Chiorescu}\ \emph {et~al.}(2010)\citenamefont
  {Chiorescu}, \citenamefont {Groll}, \citenamefont {Bertaina}, \citenamefont
  {Mori},\ and\ \citenamefont {Miyashita}}]{PhysRevB.82.024413}%
  \BibitemOpen
  \bibfield  {author} {\bibinfo {author} {\bibfnamefont {I.}~\bibnamefont
  {Chiorescu}}, \bibinfo {author} {\bibfnamefont {N.}~\bibnamefont {Groll}},
  \bibinfo {author} {\bibfnamefont {S.}~\bibnamefont {Bertaina}}, \bibinfo
  {author} {\bibfnamefont {T.}~\bibnamefont {Mori}}, \ and\ \bibinfo {author}
  {\bibfnamefont {S.}~\bibnamefont {Miyashita}},\ }\href {\doibase
  10.1103/PhysRevB.82.024413} {\bibfield  {journal} {\bibinfo  {journal} {Phys.
  Rev. B}\ }\textbf {\bibinfo {volume} {82}},\ \bibinfo {pages} {024413}
  (\bibinfo {year} {2010})}\BibitemShut {NoStop}%
\bibitem [{\citenamefont {Auff{\`{e}}ves}\ \emph {et~al.}(2008)\citenamefont
  {Auff{\`{e}}ves}, \citenamefont {Besga}, \citenamefont {G{\'{e}}rard},\ and\
  \citenamefont {Poizat}}]{PhysRevA.77.063833}%
  \BibitemOpen
  \bibfield  {author} {\bibinfo {author} {\bibfnamefont {A.}~\bibnamefont
  {Auff{\`{e}}ves}}, \bibinfo {author} {\bibfnamefont {B.}~\bibnamefont
  {Besga}}, \bibinfo {author} {\bibfnamefont {J.-M.}\ \bibnamefont
  {G{\'{e}}rard}}, \ and\ \bibinfo {author} {\bibfnamefont {J.-P.}\
  \bibnamefont {Poizat}},\ }\href {\doibase 10.1103/PhysRevA.77.063833}
  {\bibfield  {journal} {\bibinfo  {journal} {Phys. Rev. A}\ }\textbf {\bibinfo
  {volume} {77}},\ \bibinfo {pages} {063833} (\bibinfo {year}
  {2008})}\BibitemShut {NoStop}%
\bibitem [{\citenamefont {Nikoghosyan}\ and\ \citenamefont
  {Fleischhauer}(2009)}]{PhysRevLett.103.163603}%
  \BibitemOpen
  \bibfield  {author} {\bibinfo {author} {\bibfnamefont {G.}~\bibnamefont
  {Nikoghosyan}}\ and\ \bibinfo {author} {\bibfnamefont {M.}~\bibnamefont
  {Fleischhauer}},\ }\href {\doibase 10.1103/PhysRevLett.103.163603} {\bibfield
   {journal} {\bibinfo  {journal} {Phys. Rev. Lett.}\ }\textbf {\bibinfo
  {volume} {103}},\ \bibinfo {pages} {163603} (\bibinfo {year}
  {2009})}\BibitemShut {NoStop}%
\bibitem [{\citenamefont {Sachdev}(1984)}]{PhysRevA.29.2627}%
  \BibitemOpen
  \bibfield  {author} {\bibinfo {author} {\bibfnamefont {S.}~\bibnamefont
  {Sachdev}},\ }\href {\doibase 10.1103/PhysRevA.29.2627} {\bibfield  {journal}
  {\bibinfo  {journal} {Phys. Rev. A}\ }\textbf {\bibinfo {volume} {29}},\
  \bibinfo {pages} {2627} (\bibinfo {year} {1984})}\BibitemShut {NoStop}%
\bibitem [{\citenamefont {Wesenberg}\ \emph {et~al.}()\citenamefont
  {Wesenberg}, \citenamefont {Kurucz},\ and\ \citenamefont {M{\o}lmer}}]{WKM}%
  \BibitemOpen
  \bibfield  {author} {\bibinfo {author} {\bibfnamefont {J.~H.}\ \bibnamefont
  {Wesenberg}}, \bibinfo {author} {\bibfnamefont {Z.}~\bibnamefont {Kurucz}}, \
  and\ \bibinfo {author} {\bibfnamefont {K.}~\bibnamefont {M{\o}lmer}},\
  }\href@noop {} {\enquote {\bibinfo {title} {Dynamics of the collective modes
  of an inhomogeneous spin ensemble in a cavity},}\ }\Eprint
  {http://arxiv.org/abs/arXiv:1008.5197} {arXiv:1008.5197} \BibitemShut
  {NoStop}%
\bibitem [{\citenamefont {Kurucz}\ \emph {et~al.}(2009)\citenamefont {Kurucz},
  \citenamefont {S{\o}rensen}, \citenamefont {Taylor}, \citenamefont {Lukin},\
  and\ \citenamefont {Fleischhauer}}]{PhysRevLett.103.010502}%
  \BibitemOpen
  \bibfield  {author} {\bibinfo {author} {\bibfnamefont {Z.}~\bibnamefont
  {Kurucz}}, \bibinfo {author} {\bibfnamefont {M.~W.}\ \bibnamefont
  {S{\o}rensen}}, \bibinfo {author} {\bibfnamefont {J.~M.}\ \bibnamefont
  {Taylor}}, \bibinfo {author} {\bibfnamefont {M.~D.}\ \bibnamefont {Lukin}}, \
  and\ \bibinfo {author} {\bibfnamefont {M.}~\bibnamefont {Fleischhauer}},\
  }\href {\doibase 10.1103/PhysRevLett.103.010502} {\bibfield  {journal}
  {\bibinfo  {journal} {Phys. Rev. Lett.}\ }\textbf {\bibinfo {volume} {103}},\
  \bibinfo {pages} {010502} (\bibinfo {year} {2009})}\BibitemShut {NoStop}%
\end{thebibliography}%
\end{document}